\documentclass[apj]{emulateapj}
\usepackage{apjfonts}
\usepackage{amsmath}
\usepackage{graphicx}
\usepackage{natbib}
\usepackage[usenames,dvips]{color}
\shorttitle{Energy Dissipation through Quasi-Static Tides in White Dwarf Binaries}
\shortauthors{Willems, Deloye, \& Kalogera}

\begin{document}

\title{Energy Dissipation through Quasi-Static Tides in White Dwarf Binaries}

\author{B.\ Willems, C.J.\ Deloye, V.\ Kalogera}
\affil{Northwestern University, Department of Physics and Astronomy, 2131
  Tech Drive, Evanston, IL 60208, USA}
\email{b-willems@northwestern.edu, cjdeloye@northwestern.edu, vicky@northwestern.edu}

\begin{abstract}
We present a formalism to study tidal interactions in white dwarf binaries in the limiting case of 
quasi-static tides, in which the tidal forcing frequencies are small compared to the inverse of the 
white dwarf's dynamical time scale. The formalism is valid for arbitrary orbital eccentricities and 
therefore applicable to white dwarf binaries in the Galactic disk as well as globular clusters. In the 
quasi-static limit, the total perturbation of the gravitational potential shows a phase shift with respect 
to the position of the companion, the magnitude of which is determined primarily by the efficiency 
of energy dissipation through convective damping. We determine rates of secular evolution of the 
orbital elements and white dwarf rotational angular velocity for a $0.3\,M_\odot$ helium white 
dwarf in binaries with orbital frequencies in the Laser Interferometer Space Antenna (LISA) gravitational 
wave frequency band and companion masses ranging from $0.3\,M_\odot$ to $10^5\,M_\odot$. The
resulting tidal evolution time scales for the orbital semi-major axis are longer than a Hubble time, 
so that convective damping of quasi-static tides need not be considered in the construction of
gravitational wave templates of white dwarf binaries in the LISA band. Spin-up of the white dwarf, 
on the other hand, can occur on time scales of less than 10\,Myr, provided that the white dwarf is 
initially rotating with a frequency much smaller than the orbital frequency. For semi-detached white dwarf 
binaries spin-up can occur on time scales of less than 1\,Myr. Nevertheless, the time scales 
remain longer than the orbital inspiral time scales due to gravitational radiation, so that the degree of
asynchronism in these binaries increases. As a consequence, tidal forcing eventually occurs at
forcing frequencies beyond the quasi-static tide approximation. For the shortest period binaries, energy 
dissipation is therefore expected to take place through dynamic tides and resonantly excited $g$-modes. 
\end{abstract}


\section{Introduction}

White dwarfs are the most common endpoint of stellar evolution in galaxies and 
dense stellar systems. 
During the past decade, ongoing optical surveys such as the Sloan Digital Sky Survey (SDSS) and the ESO 
Supernova Ia Progenitor surveY (SPY) have provided a plethora of new white dwarf binaries suitable to 
study poorly understood binary evolution phases such as common envelope evolution and type Ia supernovae.  
Binaries consisting of two white dwarfs are furthermore the single most abundant and only 
guaranteed sources of gravitational wave radiation in the Galaxy for the Laser Interferometer Space Antenna,  
LISA (Bender et al. 1998). 

Despite the high abundance of white dwarf binaries in present-day and planned astrophysical surveys, the 
study of their orbital evolution has remained limited to models in which the white dwarfs are treated as point 
masses, in so far as the calculation of the gravitational potential is concerned. Gravitational wave forms of double white dwarf binaries, for instance, currently only account for 
orbital frequency changes driven by gravitational radiation. As a significant fraction of these systems spirals in to 
periods as short as 5 minutes, tidal effects can, in principle, have a significant impact on the orbital evolution and 
thus the gravitational wave frequency evolution. The impact 
depends strongly on the strength and the nature of the tidal energy dissipation mechanism, which, for white 
dwarfs, remains uncertain. Racine, Phinney, \& Arras (2007) recently also proposed a 
non-dissipative tidal evolution mechanism in which angular momentum exchange between the star and the 
orbit is driven by resonant excitation of Rossby modes. However, as commented by the authors, non-linear effects 
likely limit the lifetime of the resonance and thus the tidal evolution mechanism. 

Campbell (1984) studied dissipative tides in white dwarf binaries assuming the orbital frequency of the 
binary to be much smaller than the frequencies of the white dwarf's free modes of oscillation and limiting 
his study to circular binaries. The author considered energy dissipation through perturbations of the radiative 
energy flux, and used a perturbation technique to calculate the tidal velocity field in a non-rotating white 
dwarf. Campbell found that, in circular binaries, the synchronization time scale of a white dwarf can be comparable to the white dwarf's lifetime, provided that the initial degree of asynchronism is sufficiently 
large. However, in his estimate for the tidal synchronization time scale, he used a rather high white dwarf 
luminosity of $0.03\,L_\odot$, which decreases the time scale by several orders of magnitude compared to 
the tidal synchronization time scale of a more conventional $\sim 10^{-5}\,L_\odot$ white dwarf (see his 
Eq. 46). Other authors studying the impact of tidal effects on the evolution of white dwarf binaries either 
assumed tidal dissipation to be strong enough to maintain synchronism at all times (Webbink \& Iben 1987; Mochkovitch \& Livio 1989; Iben, Tutukov, \& Fedorova 1998) or parameterized tidal dissipation by means 
of an ad-hoc tidal synchronization time scale (Marsh, Nelemans, \& Steeghs 2004; Gokhale, Peng, \& Frank 
2007).   

In contrast, tidal evolution theories for binaries with non-degenerate component stars are well developed 
and able to pass stringent observational tests such as measured orbital decay rates of high-mass X-ray binaries 
(Belczynski et al. 2008) and circularization cut-off periods in young open cluster binaries 
(Witte \& Savonije 2002, Zahn 2008). However, for the latter test, some debate still exists on the agreement 
between theory and observations  (Mathieu, Meibom, \& Dolan 2004; Meibom \& Mathieu 2005). 
According to our current understanding, non-degenerate stars with convective envelopes dissipate 
tidal energy primarily through convective damping 
of quasi-static tides, while non-degenerate stars with radiative envelopes dissipate tidal energy primarily 
through radiative damping of dynamic tides (Zahn 1975, 1977). Both dissipation mechanisms may be 
significantly enhanced by resonances between tidally forced oscillations and free oscillation modes of the 
component stars, especially in binaries with eccentric orbits (Witte \& Savonije 1999, 2001, 2002). 

Our aim in this paper is to initiate a systematic investigation of tidal dissipation mechanisms operating in 
white dwarfs by building on the success of tidal evolution for non-degenerate 
stars. Here we explore the effectiveness of convective damping of quasi-static tides as a tidal 
energy dissipation mechanism in white dwarf binaries. In \S\,2 and \S\,3, we outline the basic assumptions 
and introduce the system of differential equations governing tidally forced oscillations in close binaries. In 
\S\,4, we use a perturbation method to derive approximate solutions to the system of differential equations 
appropriate for quasi-static tides. In \S\,5 and \S\,6, we present the equations governing the secular evolution 
of the orbital elements and white dwarf rotation rates. In \S\,7, we calculate the quasi-static tidal distortion 
and orbital evolution time scales for a $0.3\,M_\odot$ He white dwarf model, and compare the orbital evolution 
time scales with those due to gravitational radiation. The final section is devoted to concluding remarks.

\section{The tide-generating potential}
\label{secpot}

We consider a close binary system of stars revolving around one another in a Keplerian orbit with period $P_
{\rm orb}$, semi-major axis $a$, and eccentricity $e$. The first star, with mass $M_1$, radius $R_1$, and 
luminosity $L_1$, rotates uniformly around an axis perpendicular to the  orbital plane with angular velocity 
$\vec{\Omega}_1$ in the sense of the orbital motion.  The second star, hereafter referred to as the 
companion, has mass $M_2$ and is considered to be a point mass. The rotational angular velocity $
\Omega_1$ of star~1 is assumed to be small enough for the Coriolis force and the centrifugal force to be 
negligible, so that the tides raised by the companion can be treated as forced perturbations of a non-rotating, 
spherically symmetric, equilibrium star. 

The tidal force exerted by the companion is derived from the tide-generating potential $ \varepsilon_{\rm 
tide}\, W(\vec{r},t)$ as
\begin{equation}
\vec{F}_{\rm tide} = - \varepsilon_{\rm tide} \vec{\nabla} W
\end{equation}
with
\begin{equation}
\varepsilon_{\rm tide} \equiv \left( {R_1 \over a} \right)^3 {M_2 \over M_1}.
\end{equation}
The quantity $\varepsilon_{\rm tide}$ is a small dimensionless parameter corresponding to the ratio of the 
tidal force to the gravity at the star's equator. We express the tide-generating potential in terms of spherical 
coordinates $\vec{r} = (r,\theta,\phi)$ with respect to an orthogonal frame of reference that is corotating 
with the star. The polar angle $\theta$ is measured from the rotational angular velocity vector, while the 
azimuthal angle $\phi$ is measured in the orbital plane in the sense of the orbital motion. At $t=0$, the $\phi=0$ 
direction coincides with the direction from the star's mass center to the periastron of the binary 
orbit\footnote{For binaries with circular orbits, the $\phi=0$ direction coincides 
with the direction from the star's mass center to the ascending node of the binary orbit at $t=0$.}. 

As is customary, we expand the tide-generating potential in Fourier series as
\begin{eqnarray}
\lefteqn{\varepsilon_{\rm tide}\, W \left( \vec{r},t \right) = - \varepsilon_{\rm tide}\, {{G\,M_1} \over 
R_1}\, 
  \sum_{\ell=2}^4 \sum_{m=-\ell}^\ell \sum_{k=-\infty}^\infty 
  c_{\ell,m,k} }  \nonumber \\
 & & \phantom{ttt} \times \left({r\over R_1}\right)^\ell\,
  Y_\ell^m(\theta,\phi)\, 
  \exp \left[{\rm i} \left(\sigma_{m,k}\, t
  - k\, n\, \tau \right)\right]  \label{W} 
\end{eqnarray}
(e.g. Polfliet \& Smeyers 1990). In this expansion, $G$ is the Newtonian constant of gravitation, $Y_\ell^m
(\theta,\phi)$ an unnormalized spherical harmonic of degree $\ell$ and azimuthal number $m$, $\sigma_
{m,k} = k\, n + m\, \Omega_1$ a forcing angular frequency with respect to the corotating frame of 
reference, $n=2\,\pi/P_{\rm orb}$ the mean motion, and $\tau$ a time of periastron passage. The factors $c_
{\ell,m,k}$ are Fourier coefficients defined as
\begin{eqnarray}
\lefteqn{c_{\ell,m,k} = \displaystyle
  {{(\ell-|m|)!} \over {(\ell+|m|)!}}\, P_\ell^{|m|}(0)
  \left({R_1\over a}\right)^{\ell-2}
  {1\over {\left({1 - e^2}\right)^{\ell - 1/2}}} } \nonumber \\
 & & \phantom{ttt} \times {1\over \pi} {\int_0^\pi (1 + e\, \cos v)^{\ell-1}\,
  \cos (k\, M + m\, v)\, dv}, \label{clmk}
\end{eqnarray}
where $v$ is the true anomaly, $M=n(t-\tau)$ the mean anomaly, and $P_\ell^m(x)$ an associated Legendre 
polynomial of the first kind. The coefficients $c_{\ell,m,k}$ obey the property of symmetry $c_{\ell,-m,-k} 
= c_{\ell,m,k}$ and, since $P_\ell^{|m|}(0)=0$ for odd values of $\ell+|m|$, are equal to zero for odd 
values of $\ell+|m|$. From the binomial theorem it furthermore follows that $c_{\ell,m,0} = 0$ for $m = 
\pm \ell$. For a given orbital eccentricity and sufficiently large Fourier indices $k$, the non-zero 
coefficients $c_{\ell,m,k}$ decrease with increasing values of $k$, though the decrease is slower for higher 
orbital eccentricities\footnote{For some $\ell$ and $m$ values the coefficients $c_{\ell,m,k}$ increase to a 
local maximum before decreasing with increasing values of $k$.} (e.g. Smeyers et al. 1998, Willems 2003). 
There is thus only a finite number of non-negligible terms contributing to the expansion of the tide-
generating potential, and the number of non-negligible terms increases with increasing orbital eccentricities.

The expansion of the tide-generating potential in Fourier series introduces an infinite number of forcing 
angular frequencies $\sigma_{m,k}$ in the primary. In general, the frequencies are different from zero so 
that the associated terms give rise to {\em dynamic} tides. Terms in the tide-generating potential for which $
\sigma_{m,k}=0$, on the other hand, give rise to {\em
static} tides. We note that at least one such static tide exists for each spherical harmonic degree $\ell$: the 
tide generated by the term associated with $k = m = 0$. In binaries with spin-orbit resonances, additional 
static tides exist for non-zero values of $k$ and $m$ satisfying $k/m = - \Omega_1/n$.  

From the definition of the Fourier coefficients $c_{\ell,m,k}$ it follows that the $\ell=3$ and $\ell=4$ terms 
in the expansion of the tide-generating potential contain additional factors $R_1/a$ in comparison to the $
\ell=2$ terms. The expansion of the tide-generating potential is therefore usually restricted to the dominant $
\ell=2$ terms. However, in the case of binaries in which the primary is close to or is filling its Roche lobe  
(e.g.\ inspiralling double white dwarfs and AM\,CVn binaries), the ratio $R_1/a$ may become large enough 
so that the $\ell=3$ and $\ell=4$ terms are no longer negligible. 

In the particular case of a binary with a circular orbit, the Fourier coefficients $c_{\ell,m,k}$ are different 
from zero only when $k=-m$. The non-zero coefficients are then given by
\begin{equation}
c_{\ell,m,-m} = \displaystyle
  {{(\ell-|m|)!} \over {(\ell+|m|)!}}\, P_\ell^{|m|}(0)
  \left({R_1\over a}\right)^{\ell-2}, \label{clmk2}
\end{equation}
and the tide-generating potential reduces to
\begin{eqnarray}
\lefteqn{ \varepsilon_{\rm tide}\, W \left( \vec{r}, t \right)
  = - \varepsilon_{\rm tide}\, {{G\,M_1} \over R_1}\,
  \sum_{\ell=2}^4 \sum_{m=-\ell}^\ell  
  c_{\ell,m,-m}\, 
  } \nonumber \\
 & & \phantom{ttt} \times \left({r\over R_1}\right)^\ell\, Y_\ell^m(\theta,\phi)
  \exp \left[{\rm i}\,m \left( \Omega_1\,t - M \right)\right].
  \label{pote0}
\end{eqnarray}
Hence, for binaries with circular orbits, the expansion of the tide-generating potential becomes independent 
of time with respect to the co-rotating frame of reference when the star's rotation rate is synchronized with 
the orbital motion. The tide-generating potential then corresponds to the potential giving rise to {\em 
equilibrium} tides. The values of the coefficients $c_{\ell,m,k}$ for binaries with circular orbits are listed in 
Table~\ref{clmkcirc}. 

\begin{deluxetable*}{lccccc}
\tablewidth{12.0cm}
\tablecolumns{6}
\tablecaption{Fourier coefficients $c_{\ell,m,-m}$ for binaries with circular orbits. \label{clmkcirc}}
\tablehead{
    & \colhead{$m=0$} & \colhead{$m= \pm 1$} & \colhead{$m= \pm 2$} & \colhead{$m= \pm 3$} & 
\colhead{$m= \pm 4$}  
   }
\startdata
$\ell=2$ & -1/2 &  0  & 1/8   &  -  &  -   \\
$\ell=3$ &  0  & $(1/8)(R_1/a)$ &  0  & $-(1/48)(R_1/a)$ &  -   \\
$\ell=4$ &  $(3/8)(R_1/a)^2$  &  0  & $-(1/48)(R_1/a)^2$ &  0  & $(1/348)(R_1/a)^2$ 
\enddata
\end{deluxetable*}

\begin{figure}
\begin{center}
\resizebox{8.0cm}{!}{\includegraphics{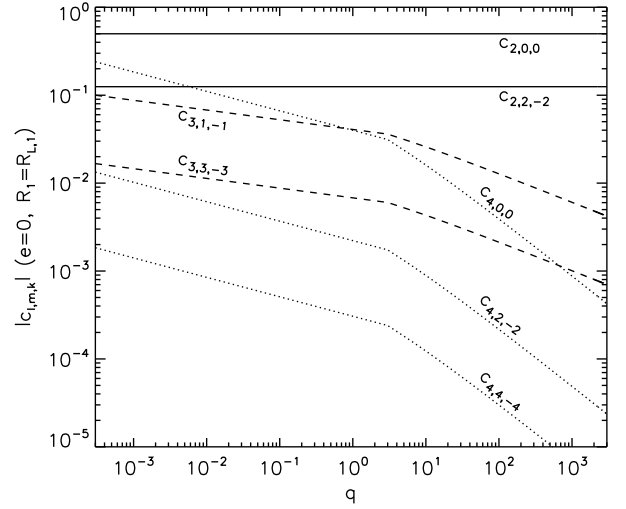}}
\end{center}
\caption{Absolute value of the non-zero Fourier coefficients $c_{\ell,m,-m}$ as functions of the mass ratio 
$q = M_2/M_1$, for circular binaries with a Roche-lobe filling primary. Since $c_{\ell,m,-m} = c_{\ell,-
m,m}$, only Fourier coefficients with positive values of $m$ are shown.}
\label{clmkfig2}
\end{figure}

If the primary in a binary with a circular orbit fills its Roche lobe, its radius $R_1$ is approximately equal 
to the volume-equivalent radius $R_{\rm L,1}$ of its Roche lobe. In terms of the mass ratio $q=M_2/M_1$, 
the latter can be expressed as (Eggleton 1983) 
\begin{equation}
{R_{\rm L,1} \over a} = {{0.49\, q^{-2/3}} \over {0.6\, q^{-2/3} + \ln \left(
  1 + q^{-1/3} \right)}},  \label{RL}
\end{equation}
so that the Fourier coefficients $c_{\ell,m,k}$ become a function of the mass ratio $q$. The variations of 
these coefficients as functions of $q$ are shown in Fig.~\ref{clmkfig2}. For large mass ratios $q\gg 1$, the 
second-degree Fourier coefficients $c_{2,m,k}$ are always at least an order of magnitude larger than the 
Fourier coefficients $c_{3,m,k}$ and $c_{4,m,k}$, so that the $\ell=3$ and $\ell=4$ terms can be neglected 
in the expansion of the tide-generating potential. However, when $q \lesssim 1$, the magnitude of the 
Fourier coefficients $c_{3,1,-1}$, $c_{3,-1,1}$, and $c_{4,0,0}$ becomes comparable to that of the 
coefficients $c_{2,2,-2}$ and $c_{2,-2,2}$. The other non-zero Fourier coefficients remain at least an order 
of magnitude smaller than $c_{2,2,-2}$ and $c_{2,-2,2}$. Since non-Roche-lobe filling primaries always 
have $R_1 < R_{\rm L,1}$, the  curves shown in Fig.~\ref{clmkfig2} pose upper limits on the Fourier 
coefficients $c_{\ell,m,k}$ in detached binaries with circular orbits. 

In the following sections, we derive a formalism to study quasi-static tides in close binaries. 
Given the potentially non-trivial role of the $\ell=3$ and $\ell=4$ terms, we allow for arbitrary values
of $\ell$, though in the applications we restrict ourselves to mass ratios $q \ge 1$ for which it is 
sufficient to only consider terms in the tide-generating potential associated with $\ell=2$.

\section{The equations governing forced oscillations of a spherically symmetric star}
\label{goveqs}

When the tidal force exerted by the companion is treated as a small perturbing force acting on a static, 
spherically symmetric, equilibrium star, the equations governing the tidal displacement field $\vec{\xi}_
{\rm tide}$ and the perturbations of the star's mass density $\rho$, pressure $P$, and potential of self-
gravity $\Phi$ are obtained by perturbing and linearizing the  equation of motion, the equation expressing 
the conservation of mass, the energy equation, and the Poisson equation. The perturbed and linearized 
equations take the form (e.g. Ledoux \& Walraven 1958)
\begin{eqnarray}
\lefteqn{ {{\partial^2 \vec{\xi}_{\rm tide}} \over {\partial t^2}} =  
  - \nabla \Psi_{\rm tide} - {1 \over \rho}\, \nabla P_{\rm tide}^\prime 
  + {1 \over \rho^2}\, {{dP} \over {dr}}\, \rho_{\rm tide}^\prime } \label{eq1} \\
\lefteqn{{\rho_{\rm tide}^\prime \over \rho} + {1 \over \rho}\,
  {{d\rho} \over {dr}} \xi_r 
  = - \nabla \cdot \vec{\xi}_{\rm tide}, } \label{eq2} \\
\lefteqn{{\partial \over {\partial t}} \left\{
  {P_{\rm tide}^\prime \over P} + {1 \over P}\,
  {{dP} \over {dr}}\, \xi_r - \Gamma_1 
  \left[ {\rho_{\rm tide}^\prime \over \rho} + {1 \over \rho}\,
  {{d\rho} \over {dr}}\, \xi_r 
  \right] \right\} } \nonumber \\
 & & \phantom{www} = {{\left( \Gamma_3 - 1 \right) \rho} \over P}  
  \left( {{dQ}\over{dt}}\right)_{\rm tide}^{\! \prime}, 
   \hspace{4cm} \label{eq3} \\
\lefteqn{\nabla^2 \Phi_{\rm tide}^\prime = 4\, \pi\, G\, 
  \rho_{\rm tide}^\prime,} 
  \label{eq4}
\end{eqnarray}
where $\Psi_{\rm tide} = \Phi_{\rm tide}^\prime + \varepsilon_{\rm tide}\, W$ is the total perturbation of 
the gravitational potential, $\xi_r$ the radial component of the tidal displacement field, $dQ/dt$ the rate of 
change of thermal energy (see Appendix~\ref{thermal} for details), and a prime on a quantity denotes the Eulerian perturbation of that quantity. The 
generalized isentropic coefficients $\Gamma_1$ and $\Gamma_3$ are defined as 
\begin{equation}
\Gamma_1 = [(\partial \ln P)/(\partial \ln \rho)]_S,
\end{equation}
and
\begin{equation} 
\Gamma_3-1 = [(\partial \ln T)/(\partial \ln \rho)]_S,
\end{equation} 
where $S$ is the entropy, and $T$ the temperature (e.g., Cox \& Giuli 1968). 

Following Savonije \& Witte (2002), we furthermore account for the effects of turbulent convection on the 
tides raised by the companion by adding a radial viscous force density
\begin{equation}
f_{r, {\rm visc}} = {1 \over {\rho\,r^2}}\, {\partial \over {\partial r}}
  \left[ \rho\, r^2\, \nu {\partial \over {\partial r}} \left( {{\partial  \xi_{r,{\rm tide}}} \over {\partial t}} 
  \right) \right]
\end{equation}
to the right-hand member of Eq.~(\ref{eq1}). In this equation, $\nu$ is the coefficient of turbulent 
viscosity. The viscous force is different from zero only in convective regions of the star. As outlined
in Appendix~\ref{thermal}, we do not incorporate effects of convection on the perturbation of the rate of 
change of thermal energy. 

Next, we separate the time and angular coordinates in Eqs.~(\ref{eq1})--(\ref{eq4}) by expanding the tidal 
displacement field as
\begin{eqnarray}
\lefteqn{ \vec{\xi}_{\rm tide} \left( \vec{r}, t \right) = 
  \displaystyle \sum_{\ell=2}^4
  \sum_{m=-\ell}^\ell \sum_{k=-\infty}^\infty 
  \left[ \xi_{\ell,m,k}(r), {{\eta_{\ell,m,k}(r)} \over r}
  {\partial \over{\partial \theta}}, {{\eta_{\ell,m,k}(r)} \over {r\, \sin \theta}}
  {\partial \over{\partial \phi}} \right] } \nonumber \\
 & & \times \, Y_\ell^m (\theta,\phi)\, \exp \left[ {\rm i} \left( \sigma_{m,k}\, t 
  - k\, n\, \tau \right) \right],  \hspace{3.7cm}  \label{exp1}  
\end{eqnarray}
and the perturbations of the stellar structure quantities as
\begin{eqnarray}
\lefteqn{f^\prime_{\rm tide} \left( \vec{r}, t \right) = 
  \displaystyle \sum_{\ell=2}^4
  \sum_{m=-\ell}^\ell \sum_{k=-\infty}^\infty 
  f_{\ell,m,k}^\prime(r) \, Y_\ell^m (\theta,\phi) } \nonumber \\
 & & \phantom{ww} \times \, \exp \left[ {\rm i} \left( \sigma_{m,k}\, t 
  - k\, n\, \tau \right) \right].  \hspace{3.0cm}  \label{fprime}
\end{eqnarray}
Substitution of the expansions in Eqs.~(\ref{eq1})--(\ref{eq4}) and replacing the $\theta$ and $\phi$ 
components of the equation of motion by the radial component of the vorticity equation and the equation for 
the divergence of the transverse component of the tidal displacement field (see, e.g., Ledoux \& Walraven 
1958, Aizenman \& Smeyers 1977, Willems 2000) yields  
\begin{eqnarray}
\lefteqn{\sigma_{m,k}^2\, \xi_{\ell,m,k} 
  = {{d\Psi_{\ell,m,k}}\over{dr}} - 
  {\rho_{\ell,m,k}^\prime \over \rho^2}\, {{dP}\over{dr}} + 
  {1\over \rho}\, {{d P_{\ell,m,k}^\prime}
  \over {dr}}}  \nonumber \\
 & & \phantom{www} - {\rm i} \sigma_{m,k}\, {1 \over {\rho\, r^2}} {d \over {dr}} \left( \rho\, r^2\,
   \nu\, {{d\xi_{\ell,m,k}} \over {dr}} \right), \hspace{2.4cm}  \label{sep1a} 
\end{eqnarray}
\begin{equation}
\sigma_{\ell,m,k}^2\, \eta_{\ell,m,k} = 
  \Psi_{\ell,m,k} + {P_{\ell,m,k}^\prime \over \rho},  \hspace{3.2cm} 
  \label{sep1b} 
\end{equation}
\begin{equation}
{\rho_{\ell,m,k}^\prime \over \rho} + {1 \over \rho}\,
  {{d\rho} \over {dr}}\, \xi_{\ell,m,k}
  = - \alpha_{\ell,m,k},  \hspace{3.1cm}  \label{sep4} 
\end{equation}
\begin{eqnarray}
\lefteqn{{\rm i}\, \sigma_{m,k} \left(
  {P_{\ell,m,k}^\prime \over P} + {1 \over P}\,
  {{dP} \over {dr}}\, \xi_{\ell,m,k} + \Gamma_1\, 
  \alpha_{\ell,m,k} \right)} \nonumber \\
 & & \phantom{wwww} = {{\left( \Gamma_3 - 1 \right) \rho} \over P}\,  
  \left( {{dQ}\over{dt}}\right)_{\ell,m,k}^\prime, 
  \hspace{2.9cm} \label{sep5} 
\end{eqnarray}
\begin{equation}
{1 \over r^2}\, {d \over {dr}} \left( r^2 
  {{d \Psi_{\ell,m,k}} \over {dr}} \right) - {{\ell(\ell+1)}
  \over r^2}\, \Psi_{\ell,m,k} = 4\, \pi\, G\, 
  \rho_{\ell,m,k}^\prime.  \hspace{0.1cm}  \label{sep6}
\end{equation}
Here, $\alpha_{\ell,m,k}$ is the divergence of the tidal displacement field given by 
\begin{equation}
\alpha_{\ell,m,k} = {1 \over r^2}\, {d \over {dr}} \left( r^2\, 
  \xi_{\ell,m,k} \right) - {{\ell(\ell+1)} \over r^2}\, 
  \eta_{\ell,m,k}.   \hspace{1.0cm}   \label{sep7}
\end{equation}

In the next section, we derive approximate solutions to the above system of differential equations in the 
limiting case of small forcing frequencies. For this purpose, it is convenient to use Eqs.~(\ref{sep1b}), 
(\ref{sep4}), and (\ref{sep5}) to eliminate $\Psi$, $\rho^\prime$, and $P^\prime$ from 
Eq.~(\ref{sep1a}). It follows that
\begin{eqnarray}
\lefteqn{ {N^2 \over g}\, c_s^2\, \alpha_{\ell,m,k} = 
  \sigma_{m,k}^2 \left( \xi_{\ell,m,k} - {{d\eta_{\ell,m,k}} \over {dr}} \right) }  \nonumber \\
 & & \phantom{ww} + {\rm i}\, \sigma_{m,k}\, {1 \over {\rho\, r^2}} {d \over {dr}} \left( \rho\, r^2\,
   \nu\, {{d\xi_{\ell,m,k}} \over {dr}} \right)  \nonumber \\
 & & \phantom{ww} + \frac{\rm i}{\sigma_{m,k}}\, {1 \over \rho}\, {{d\rho} \over {dr}}
  \left( \Gamma_3-1 \right) \left( {{dQ}\over{dt}}\right)_{\ell,m,k}^\prime, \hspace{1.5cm}  \label{elim0} 
\end{eqnarray}
where $N^2 = -g [(g/c_s^2) + (d\ln\rho/dr)]$ is the square of the Brunt-V\"{a}is\"{a}l\"{a} frequency,  $g$ 
the gravity, and $c_s^2 = \Gamma_1 P / \rho$ the square of the isentropic sound speed.  

Equations~(\ref{sep1a})--(\ref{sep6}) must be supplemented with boundary conditions at the star's center 
and at the star's surface. At $r=0$, we impose that the tidal displacement field remains finite. At $r=R_1$, 
we adopt zero-boundary conditions and impose the Lagrangian displacement of the pressure to vanish: $
(\delta P)_{\rm tide}(R_1) = 0$. We furthermore impose the gravitational potential and its first derivative to 
be continuous at $r=R_1$, which is expressed by the condition
\begin{eqnarray}
\lefteqn{ \left( {{d \Psi_{\ell,m,k}} \over {dr}} \right)_{R_1} + {{\ell+1} \over R_1}\, \Psi_{\ell,m,k}
(R_1) 
  + 4\, \pi\, G\, \rho(R_1)\, \xi_{\ell,m,k}(R_1) }  \nonumber \\
 & & \phantom{www}  = - \varepsilon_{\rm tide}\, (2\, \ell + 1) {{G\, M_1} \over R_1^2}\, c_{\ell,m,k}  
  \hspace{3cm} \label{bcpsi}
\end{eqnarray}
(e.g.\ Polfliet \& Smeyers 1990). Because of the non-homogeneous term in the right-hand member of this 
equation, the solutions to Eqs.~(\ref{sep1a})--(\ref{sep6}) are proportional to the product $\varepsilon_{\rm 
tide}\, c_{\ell,m,k}$.

A system of equations of the form of Eqs.~(\ref{sep1a})--(\ref{sep6}) exists for each $\ell$, $m$, 
and $k$ in the expansion of the tide-generating potential. The system of equations is complex due the 
presence of the convective damping term in the right-hand member of  Eq.~(\ref{sep1a}) and the 
perturbation of the rate of change of thermal energy in the right-hand member of  Eq.~(\ref{sep5}). The 
solutions will therefore show a phase shift with respect to those found in the adiabatic approximation. We 
furthermore note that the solution to Eqs.~(\ref{sep1a})--(\ref{sep6}) associated with the forcing angular 
frequency $-\sigma_{m,k}$ is the complex 
conjugate of the solution associated with the forcing angular frequency $\sigma_{m,k}$. The two solutions 
therefore have the same amplitude, but opposite phase shifts.

\section{The Quasi-Static Tide Approximation}
\label{qstide}

Smeyers (1997) and Smeyers \& Willems (1998) derived approximate solutions to Eqs.~(\ref{sep1a})--(\ref
{sep6}) for low-frequency dynamic tides valid to ${\cal O}(\sigma_{m,k}^2)$. The authors adopted the 
adiabatic approximation and found the radial component of the tidal displacement field to be described by a 
non-oscillatory term of ${\cal O}(\sigma_{m,k}^0)$ and an oscillatory term of ${\cal O}(\sigma_{m,k}^2)
$ (see also Zahn 1975). No term of ${\cal O}(\sigma_{m,k})$ occurs in the adiabatic approximation. 
Because of the oscillatory nature of the solutions at order ${\cal O}(\sigma_{m,k}^2)$, a multi-variable 
perturbation method was adopted to account for the large second derivatives associated with the rapid 
variations of the radial component of the tidal displacement as a function of the radial coordinate $r$.

When convective damping and the perturbation of the rate of change of thermal energy are taken into 
account, terms of ${\cal O}(\sigma_{m,k})$ appear in the low-frequency approximations of the solutions to 
Eqs.~(\ref{sep1a})--(\ref{sep6}). We therefore look for solutions valid to ${\cal O}(\sigma_{m,k})$ in the 
limiting case where the forcing angular frequency is small enough to treat the tides as quasi-static in the 
frame of reference co-rotating with the star. In this limiting case, the solutions do not show rapid variations 
in the radial direction, so that no multi-variable perturbation method is required to derive the approximate 
solutions. Our approach is similar in nature as that of Campbell (1984), except that we incorporate the
effects of convective damping and allow for arbitrary orbital eccentricities. We also make the derivation more
transparent by using the total perturbation of the gravitational potential (which is of ${\cal O}(\sigma_{m,k}^0)$) 
rather than the tidal velocity field (which is of ${\cal O}(\sigma_{m,k})$) as the main dependent variable. 

\begin{deluxetable}{llcll}
\tablecolumns{5}
\tablecaption{Units of physical quantities \label{units}}
\tablehead{\colhead{Quantity} & \colhead{Unit} & \phantom{www} & \colhead{Quantity} & \colhead
{Unit} }
\startdata
$t$ & $\left(R_1^3/G\, M_1 \right)^{1/2}$ & & $r$ & $R_1$ \\
$\xi$ & $R_1$ & & $\eta$ & $R_1^2$ \\ 
$\rho$ & $M_1/\left(4\, \pi\, R_1^3\right)$ & & $P$ & $G\, M_1^2/\left(4\, \pi\, R_1^4\right)$ \\ 
$\Phi$ & $G\, M_1/ R_1$ & & $\varepsilon_{\rm tide}\, W$ & $G\, M_1/ R_1$ \\ 
$\nu$ & $(G\, M_1\, R_1)^{1/2}$ & & $dQ/dt$ & $L_1/M_1$
\enddata
\end{deluxetable}

To derive approximate solutions to the system of equations governing tides in close binary components in 
the limiting case where $|\sigma_{m,k}| \ll (G\, M_1/R_1^3)^{1/2}$, it is convenient to pass on to 
dimensionless quantities by expressing the physical quantities in the units listed in Table~\ref{units}.
After passing on to dimensionless quantities, Eqs.~(\ref{sep5}) and~(\ref{elim0}) can be written in the form
\begin{eqnarray}
\lefteqn{ {N^2 \over g}\, c_s^2\, \alpha_{\ell,m,k} = 
  \sigma_{m,k}^2 \left( \xi_{\ell,m,k} - {{d\eta_{\ell,m,k}} \over {dr}} \right)   
  }  \nonumber \\
 & & \phantom{ww} + {\rm i} \sigma_{m,k}\, {1 \over {\rho\, r^2}} {d \over {dr}} \left( \rho\, r^2\,
   \nu\, {{d\xi_{\ell,m,k}} \over {dr}} \right) \nonumber  \\
 & & \phantom{ww} + {\rm i}\, {C \over \sigma_{m,k}}\, {1 \over \rho}\, {{d\rho} \over {dr}}
  \left( \Gamma_3-1 \right) \left( {{dQ}\over{dt}}\right)_{\ell,m,k}^\prime, \hspace{2.0cm}  \label{elim} 
\end{eqnarray}
and
\begin{eqnarray}
\lefteqn{ {P^\prime_{\ell,m,k} \over P} + {1 \over P}\, {{dP} \over {dr}}\, \xi_{\ell,m,k} + \Gamma_1\, 
  \alpha_{\ell,m,k} }  \nonumber \\
 & & \phantom{ww} = - {\rm i}\, {C \over \sigma_{m,k}}\, {{\left( \Gamma_3 - 1 \right) \rho} \over P} 
  \left( {{dQ}\over{dt}}\right)^\prime_{\ell,m,k}.  \label{sep5dim} 
\end{eqnarray}
In these equations, $C$ is the ratio of the star's dynamic time scale to its Helmholtz-Kelvin time scale: 
\begin{equation}
C = \left( {R_1^3 \over {G\,M_1}} \right)^{1/2} / \left( {{G\, M_1^2} \over {R_1\, L_1}} \right).  \label
{tdthk}
\end{equation}
In general, $C \ll \sigma_{m,k}$ so that  the terms containing the Eulerian perturbation of the rate of 
change of thermal energy in the right-hand member of Eqs.~(\ref{elim}) and~(\ref{sep5dim}) are 
non-negligible only near the star's surface where the factors $(1/\rho)(d\rho/dr)$ and $\rho/P$ become large. We 
therefore assume these terms to be of ${\cal O}(\sigma_{m,k})$ at least in some region near the star's 
surface and, following Willems et al. (2003), set 
\begin{equation}
{C \over \sigma_{m,k}} = C^\prime\, \sigma_{m,k}, \phantom{ww} C^\prime \in \mathbb{R}.  \label{C}
\end{equation}

Next, we expand the components of the tidal displacement field and the perturbed stellar structure quantities 
in series of the form
\begin{equation}
f_{\ell,m,k}(r) = f_{\ell,m,k}^{(0)}(r) + \sigma_{m,k}\, f_{\ell,m,k}^{(1)}(r) 
   + \ldots.  \label{fexp}
\end{equation}
For brevity, we  omit the subscripts $\ell$, $m$, and $k$ from the components of the tidal displacement 
field and the perturbed stellar structure quantities for the remainder of this section.

After substitution of the expansions for the components of the tidal displacement field and the perturbed 
stellar structure quantities,  Eq.~(\ref{elim}), at ${\cal O}(\sigma^0)$, yields
\begin{equation}
\alpha^{(0)} = 0
\end{equation}
in regions of the star where $N^2 \ne 0$. From Eqs.~(\ref{sep1b}), (\ref{sep4}), and~(\ref{sep5dim}), it 
then follows that  
\begin{equation}
\renewcommand{\arraystretch}{2.5}
\left.
\begin{array}{lcl}
\Psi^{(0)} = \displaystyle - {P^{\prime (0)} \over \rho}, & \phantom{....}  &
\rho^{\prime (0)} = \displaystyle - {{d\rho} \over {dr}}\, \xi^{(0)},  \\
P^{\prime (0)} = \displaystyle - {{dP} \over {dr}}\, \xi^{(0)} & \phantom{....}  &   
\xi^{(0)} = \displaystyle - {\Psi^{(0)} \over g}.
\end{array} \right\} \label{zero}
\end{equation}

Substituting these equations into Eq.~(\ref{sep6}) and making use of the equation of hydrostatic 
equilibrium, we derive a second-order differential equation for $\Psi^{(0)}$ given by
\begin{equation}
{{d^2 \Psi^{(0)}} \over {dr^2}} + {2 \over r}\, 
  {{d \Psi^{(0)}} \over {dr}} - \left[ {1 \over g}\,
  {{d\rho} \over {dr}} + {{\ell(\ell+1) } \over r^2} 
  \right] \Psi^{(0)} = 0.  \label{psi0}
\end{equation}
This equation is equivalent to the equation of Clairaut for the radial component of the tidal displacement 
field usually derived in the framework of the theory of equilibrium tides (Sterne 1939). At ${\cal O}
(\sigma^0)$, the boundary condition given by Eq.~(\ref{bcpsi}) takes the form
\begin{equation}
\left( {{d \Psi^{(0)}} \over {dr}} \right)_{r=1} 
  + \left( \ell + 1 - \frac{\rho_s}{g_s} \right)\, \Psi^{(0)}_{r=1} 
   = - \varepsilon_{\rm tide}\, (2\, \ell + 1)\, c_{\ell,m,k},    \label{bcpsi0}
\end{equation}
where $\rho_s$ and $g_s$ are the mass density and gravity at the surface of the unperturbed star, 
respectively. The solution to Eq.~(\ref{psi0}) that remains finite at the star's center and satisfies the surface 
boundary condition is real and proportional to the product $\varepsilon_{\rm tide}\, c_{\ell,m,k}$. 

At ${\cal O}(\sigma)$, Eq.~(\ref{elim}) for the divergence of the tidal displacement field yields
\begin{eqnarray}
\lefteqn{\alpha^{(1)} = 
  {\rm i}\, {g \over {c_s^2\, N^2}}\, {1 \over {\rho\, r^2}}\, {d \over {dr}} \left( \rho\, r^2\, \nu
  {{d \xi^{(0)}} \over {dr}} \right) }  \nonumber \\
 & & \phantom{ww} + {\rm i}\, C^\prime\, \left( \Gamma_3 - 1 \right) {g \over {c_s^2\, N^2}}\, {1 \over 
\rho}\,
  {{d\rho} \over {dr}}\, \left( {{dQ} \over {dt}} \right)^{\! \prime\, (0)}  \hspace{1.5cm}
\end{eqnarray}
in regions of the star where $N^2 \ne 0$. After substitution of this solution into Eqs.~(\ref{sep1b}), (\ref
{sep4}), and~(\ref{sep5dim}), we find expressions for the total perturbation of the gravitational potential, 
the Eulerian perturbations of the mass density and pressure, and the radial component of the tidal 
displacement field given by
\begin{equation}
\Psi^{(1)} = - {P^{\prime (1)} \over \rho},  \label{f1}
\end{equation}
\begin{equation}
\rho^{\prime (1)} = - {{d\rho} \over {dr}}\, \xi^{(1)} - \rho\, \alpha^{(1)},  
\end{equation}
\begin{equation}
P^{\prime (1)} = - {{dP} \over {dr}}\, \xi^{(1)} - \rho\, c_s^2\, \alpha^{(1)}
  - {\rm i}\, C^\prime\, \left( \Gamma_3 - 1 \right) \rho \left( {{dQ} \over {dt}} \right)^{\! \prime\, (0)}, 
\end{equation}
\begin{eqnarray}
\lefteqn{ \xi^{(1)} = - {\Psi^{(1)} \over g} +  {\rm i}\, 
  {1 \over {\rho\, r^2\, N^2}}\, {d \over {dr}} \left( \rho\, r^2\, \nu
  {{d \xi^{(0)}} \over {dr}} \right) }  \nonumber \\
 & & \phantom{ww} - {\rm i}\, C^\prime\, \left( \Gamma_3 - 1 \right) 
  {g \over {c_s^2\, N^2}}\, \left( {{dQ} \over {dt}} \right)^{\! \prime\, (0)}.  \label{f2}
  \hspace{2cm}
\end{eqnarray}

Finally, proceeding in a similar way as for the derivation of Eq.~(\ref{psi0}), leads to a second-order 
differential equation for $\Psi^{(1)}$:
\begin{eqnarray}
\lefteqn{{{d^2 \Psi^{(1)}} \over {dr^2}} + {2 \over r}\, 
  {{d \Psi^{(1)}} \over {dr}} - \left[ {1 \over g}\,
  {{d\rho} \over {dr}} + {{\ell(\ell+1) } \over r^2} 
  \right] \Psi^{(1)}}  \nonumber \\
 & & \phantom{ww}  = {\rm i}\,  {1 \over {g\, r^2}}\, {d \over {dr}} \left( \rho\, r^2\, \nu
 {{d \xi^{(0)}} \over {dr}} \right).   \hspace{1.5cm}  \label{psi1}
\end{eqnarray}
The associated boundary condition expressing the continuity of the gravitational potential and its first 
derivative takes the form
\begin{eqnarray}
\lefteqn{ \left( {{d \Psi^{(1)}} \over {dr}} \right)_{r=1} 
  + \left( \ell + 1 - \frac{\rho_s}{g_s} \right)\, \Psi^{(1)}_{r=1} }  \nonumber \\
 & & \phantom{www}  =  {\rm i}\, \left\{ {1 \over {N^2\, r^2}}\, {d \over {dr}}\, 
  \left[ \rho\, r^2\, \nu\, {d \over {dr}} \left( {\Psi^{(0)} \over g} \right) \right] \right\}_{r=1} 
  \nonumber \\
 & & \phantom{www}  + {\rm i}\, C^\prime \left[ \left( \Gamma_3 - 1 \right) {{\rho\, g} \over {c_s^2\, 
N^2}} 
  \left( {{dQ} \over {dt}} \right)^{\! \prime\, (0)} \right]_{r=1}.  
  \hspace{1cm} \label{bcpsi1}
\end{eqnarray}
At ${\cal O}(\sigma)$, the differential equation and surface boundary condition for the total perturbation of 
the gravitational potential are both complex. The solution therefore shows a phase shift with respect to the 
tide-generating potential, and thus also with respect to the instantaneous position of the companion. The total 
perturbation of the gravitational potential furthermore depends on the perturbation of the rate of change of 
thermal energy only through the boundary condition expressing the continuity of the gravitational potential 
and its first derivative. In Appendix~\ref{thermal}, we show that the lowest-order perturbation of the rate of 
change of thermal energy, $(dQ/dt)^{\prime (0)}$, is determined by the lowest-order solution for the total 
perturbation of the gravitational potential, $\Psi^{(0)}$, and its first derivative, $d\Psi^{(0)}\!/dr$. The 
solutions at ${\cal O}(\sigma)$ are therefore also proportional to the product $\varepsilon_{\rm tide}\, 
c_{\ell,m,k}$.  A semi-analytical solution method for the system of differential equations composed of 
Eqs.~(\ref{psi0}) and~(\ref{psi1}) and their associated boundary conditions is outlined in 
Appendix~\ref{solution}. From the solution, it follows that $\Psi^{(1)}(r)$ is purely imaginary.

\section{Orbital Evolution}

The tidal distortion of a star perturbs the spherical symmetry of it's external gravitational field, which in turn 
perturbs the motion of the companion from a pure Keplerian orbit. We study the perturbed motion in the 
framework of the theory of osculating elements in celestial mechanics (e.g. Sterne 1960, Brouwer \& 
Clemence 1961, Fitzpatrick 1970). In this framework, the rates of change of the orbital semi-major axis and 
eccentricity due a star's tidal distortion are given by
\begin{equation}
{{da} \over {dt}} = - {2 \over {n^2\,a}}\, {{\partial {\cal R}} 
  \over {\partial \tau}},  \label{da}
\end{equation}
\begin{equation}
{{de} \over {dt}} = - {1 \over {n\, a^2\, e}}\, \left[ {{1-e^2} \over n}\, 
  {{\partial {\cal R}} \over {\partial \tau}} + \left( 1 - e^2 \right)^{1/2} {{\partial {\cal R}} 
  \over {\partial \varpi}} \right],  \label{de}
\end{equation}
were $\varpi$ is the longitude of the periastron, and ${\cal R}$ is a perturbing function 
related to the Eulerian perturbation of the star's external gravitational potential $\Phi_e^\prime(\vec{r},t)$ as
\begin{equation}
{\cal R}(u,v,t) = - {{M_1 + M_2} \over M_1}\, \Phi_e^\prime \left( 
  u, {\pi \over 2}, v-\Omega_1\,t, t  \right)  \label{Ruvw}
\end{equation}
(Smeyers et al. 1991). The Eulerian perturbation of the external gravitational potential is evaluated at the 
position of the companion, $r=u$, $\theta = \pi/2$, and $\phi = v-\Omega_1\,t$, where $u$ is the distance 
between the stars.

Since ${\cal R}$ is a function of the distance $u$ and the true anomaly $v$, we transform the partial derivatives 
with respect to $\tau$ and $\varpi$ in Eqs.~(\ref{da}) and~(\ref{de}) into partial derivatives with respect to 
$u$ and $v$. By the use of the equations (e.g. Fitzpatrick 1970)
\begin{equation}
\renewcommand{\arraystretch}{1.8}
\left.
\begin{array}{l c l}
\displaystyle {{\partial u} \over {\partial \tau}} & = & \displaystyle 
  - {{n\,a\,e} \over {\left( 1-e^2 \right)^{1/2}}}\, \sin v, \\
\displaystyle {{\partial v} \over {\partial \tau}} & = & \displaystyle 
  - {n \over {\left( 1-e^2 \right)^{3/2}}}\, \left( 1 + 
  e\, \cos v \right)^2, \\  
\end{array}\right\}  \label{ddtau}
\end{equation}
the rates of change of the orbital semi-major $a$ and eccentricity $e$ due to the tidal distortion of a binary 
component then take the form
\begin{equation}
{{da} \over {dt}} = {{2\, e} \over {n\, a \left( 1-e^2 \right)^{1/2}}}
  \left[ a\, \sin v\, {{\partial {\cal R}} \over {\partial u}}
  + {{\left( 1 + e\, \cos v \right)^2} \over {e \left( 1-e^2 
  \right)}}\, {{\partial {\cal R}} \over {\partial v}} \right],  \label{da2}
\end{equation}
\begin{eqnarray}
\lefteqn{ {{de} \over {dt}} = {{\left( 1-e^2 \right)^{1/2}} \over {n\,a^2}} }  \nonumber \\
 & & \phantom{ww} \times \left\{ a\, \sin v\, {{\partial {\cal R}} \over {\partial u}} + {1 \over e}  
  \left[ {{\left( 1 + e\, \cos v \right)^2} \over {1-e^2}} - 1 \right]  
  {{\partial {\cal R}} \over {\partial v}} \right\}.  \hspace{1cm}  \label{de2}
\end{eqnarray}

The Eulerian perturbation of the external gravitational potential due to the primary's tidal distortion is a 
solution of the equation of Laplace. The solution which tends to zero at infinity can be cast in the form 
\begin{eqnarray}
\lefteqn{\Phi_e^\prime \left( \vec{r}, t \right) = 
  \sum_{\ell=0}^\infty \sum_{m=-\ell}^\ell \sum_{k=-\infty}^{+\infty} 
  A_{\ell,m,k} \left( {r \over R_1} \right)^{-(\ell+1)} 
  } \nonumber \\
 & & \phantom{ww} \times Y_\ell^m \left( \theta, \phi \right) \exp \left[ 
  {\rm i} \left( \sigma_{m,k}\,t - k\, n\, \tau 
  \right) \right], \label{Phie}
\end{eqnarray}
where the $A_{\ell,m,k}$ are constants determined by the condition that the Eulerian perturbation of the 
gravitational potential be continuous at the star's surface. 

From the definition of the total perturbation of the gravitational potential, $\Psi_{\rm tide} = \Phi_{\rm 
tide}^\prime + \varepsilon_{\rm tide}\, W$, and  Expansions~(\ref{W}) and~(\ref{fprime}), it follows that 
the Eulerian perturbation of the star's potential of self-gravity can be decomposed in terms of spherical 
harmonics and Fourier series as
\begin{eqnarray}
\lefteqn{\Phi_{\rm tide}^\prime \left( \vec{r}, t \right) = 
  \sum_{\ell=2}^\infty \sum_{m=-\ell}^\ell \sum_{k=-\infty}^{+\infty} \! \left[ \Psi_{\ell,m,k}(r)  
  + \varepsilon_{\rm tide}\, {{G\, M_1} \over R_1}\, c_{\ell,m,k} \left( {r \over R_1} \right)^\ell 
  \right] } \nonumber \\
 & & \phantom{ww} \times Y_\ell^m \left( \theta, \phi \right) \exp \left[ 
  {\rm i} \left( \sigma_{m,k}\,t - k\, n\, \tau 
  \right) \right].  \hspace{3.2cm}  \label{Phip}
\end{eqnarray}
Continuity of the gravitational potential at the star's surface thus requires
\begin{equation}
\renewcommand{\arraystretch}{1.3}
\left.
\begin{array}{lcl}
A_{\ell,m,k} = - \displaystyle 
  \varepsilon_{\rm tide}\, {{G\,M_1} \over R_1}\, 
  c_{\ell,m,k}\, 2\, F_{\ell,m,k} \,\,\, & 
  \mbox{for $\ell=2,3,4$},  \\ 
A_{\ell,m,k} = 0 \,\,\, & \,\,
  \mbox{otherwise, \phantom{ffff}} \\ 
\end{array} \right\}  \label{Almk}
\end{equation}
with
\begin{equation}
F_{\ell,m,k} = -{1 \over 2}\, \left[ {R_1 \over {G\,M_1}}\, 
  {{\Psi_{\ell,m,k}(R_1)} \over {\varepsilon_{\rm tide}\,
  c_{\ell,m,k}}} + 1 \right].  \label{Flmk}
\end{equation}
The quantities $F_{\ell,m,k}$ are dimensionless and measure the response of the star to the various forcing 
angular frequencies $\sigma_{m,k}$ appearing in the expansion of the tide-generating potential. Since $
\Psi_{\ell,m,k} \propto \varepsilon_{\rm tide}\, c_{\ell,m,k}$, the quantities are independent of the product 
$\varepsilon_{\rm tide}\, c_{\ell,m,k}$. In the limiting case of long orbital and rotational periods, the 
forcing angular frequencies $\sigma_{m,k}$ all tend to zero and the constants $F_{\ell,m,k}$ tend to the 
classical apsidal-motion constants $k_\ell$ determined in the framework of the theory of static tides 
(Smeyers \& Willems 2001). 

For tides with non-zero forcing frequencies, the quantities $F_{\ell,m,k}$ are complex. It is therefore 
convenient to write them in polar form as 
\begin{equation}
F_{\ell,m,k} = |F_{\ell,m,k}|\, \exp \left( {\rm i}\, 
  \gamma_{\ell,m,k} \right).  \label{Phi4a}
\end{equation}
Since the solution to Eqs.~(\ref{sep1a})-(\ref{sep6}) associated with the forcing angular frequency $-
\sigma_{m,k}$ has the same amplitude, but opposite phase as the solution associated with the forcing 
angular frequency $\sigma_{m,k}$, the magnitude and phase of the quantities $F_{\ell,m,k}$ obey the 
properties $|F_{\ell,-m,-k}| = |F_{\ell,m,k}|$, and $\gamma_{\ell,-m,-k} = -\gamma_{\ell,m,k}$. By means 
of these properties and the symmetry properties of the coefficients $c_{\ell,m,k}$, Expansion~(\ref{Phie}) 
for the Eulerian perturbation of the external gravitational potential can be written in real form by combining 
the terms associated with the forcing angular frequency $\sigma_{-m,-k}$ with the terms associated with the 
forcing angular frequency $\sigma_{m,k}$: 
\begin{eqnarray}
\lefteqn{{\cal R}(u,v,t) = 4\, {{G \left( M_1 + M_2 \right)} \over R_1}\,
  {M_2 \over M_1}\, \sum_{\ell=2}^4 \sum_{m=-\ell}^\ell 
  \sum_{k=0}^{+\infty} \left( {R_1 \over a} \right)^{\ell+4}
  P_\ell^{|m|}(0)
  }  \nonumber \\
 & & \times \kappa_{\ell,m,k}\, c_{\ell,m,k}\, \left|  
  F_{\ell,m,k} \right| \left( {u \over a} \right)^{-(\ell+1)}
  \cos \left( m\,v + k\,M + \gamma_{\ell,m,k} \right),  \hspace{0.8cm}  \label{R2}
\end{eqnarray}
where
\begin{equation}
\renewcommand{\arraystretch}{1.3}
\left.
\begin{array}{lll}
\kappa_{\ell,0,0} & =\, 1/2,  &  \\ 
\kappa_{\ell,m,0} & =\, 0 \,\,\,\, & 
  \mbox{for $-\ell \le m \le -1$},  \\ 
\kappa_{\ell,m,0} & =\, 1 \,\,\,\, & 
  \mbox{for \phantom{-} $1 \le m \le \ell$},  \\ 
\kappa_{\ell,m,k} & =\, 1 \,\,\,\, & 
  \mbox{for $-\ell \le m \le \ell$ and $k \ge 1$}.  \\ 
\end{array} \right\} \label{klmk}
\end{equation}

Finally, by transforming the time derivatives in Eqs.~(\ref{da2}) and~(\ref{de2}) into derivatives with 
respect to the mean anomaly and averaging over one revolution of the companion, we derive the equations 
for the rates of {\em secular} change of the orbital semi-major axis and eccentricity due to the tidal 
distortion of a binary component to be 
\begin{eqnarray}
\lefteqn{ \left( {{da} \over {dt}} \right)_{\rm sec} 
   = \frac{8\,\pi}{P_{\rm orb}}\, {M_2 \over M_1}\, a\, \sum_{\ell=2}^4 
  \sum_{m=-\ell}^\ell \sum_{k=0}^{+\infty} 
  \left( {R_1 \over a} \right)^{\ell+3}} \nonumber \\
 & & \phantom{ww} \times \kappa_{\ell,m,k}\,
  \left| F_{\ell,m,k} \right|\, \sin \gamma_{\ell,m,k}\, 
  G_{\ell,m,k}^{(2)}(e),  \hspace{3cm} \label{da4}
\end{eqnarray}
\begin{eqnarray}
\lefteqn{ \left( {{de} \over {dt}} \right)_{\rm sec} 
   = \frac{8\,\pi}{P_{\rm orb}}\, {M_2 \over M_1}\, \sum_{\ell=2}^4 
  \sum_{m=-\ell}^\ell \sum_{k=0}^{+\infty} 
  \left( {R_1 \over a} \right)^{\ell+3}} \nonumber \\
 & & \phantom{ww} \times \kappa_{\ell,m,k}\,
  \left| F_{\ell,m,k} \right|\, \sin \gamma_{\ell,m,k}\, 
  G_{\ell,m,k}^{(3)}(e),  \hspace{3cm} \label{de4}
\end{eqnarray}
where 
\begin{eqnarray}
\lefteqn{G_{\ell,m,k}^{(2)}(e) = 
  {2 \over {\left( 1-e^2 \right)^{\ell+1}}}
  c_{\ell,m,k}\, P_\ell^{|m|}(0)\,
  } \nonumber \\
 & & \times {1 \over \pi} \left[ (\ell+1)\, e \int_0^\pi
  ( 1 + e\, \cos v)^\ell\, \sin (m\,v+k\,M)\, \sin v\, dv 
  \right. \nonumber \\
 & & \left. -m \int_0^\pi (1 + e\, \cos v)^{\ell+1}
  \cos (m\,v+k\,M)\, dv \right],  \label{G2lmk}
\end{eqnarray}
\begin{eqnarray}
\lefteqn{G_{\ell,m,k}^{(3)}(e) = {1 \over {e\, \left(1-e^2\right)^{\ell}}}\,
  c_{\ell,m,k}\, P_\ell^{|m|}(0) } \nonumber \\
 & & \times {1\over \pi}\, \Biggl\{(\ell+1)\, e\, \int_0^\pi
   (1+e\,\cos v)^{\ell}\, \sin (m\,v + k\,M)\, \sin v\, dv
     \nonumber \\
 & & -\, m\, \int_0^\pi (1+e\,\cos v)^{\ell-1}\,
   \left[ (1+e\,\cos v)^2 - \left(1-e^2\right)\right]  \nonumber \\
 & & \times \cos (m\,v + k\,M)\, dv \Biggr\}.   \label{G3lmk}
\end{eqnarray}
The coefficients $G_{\ell,m,k}^{(2)}(e)$ and $G_{\ell,m,k}^{(3)}(e)$ are the same as those derived by 
Willems et al. (2003) for the rates of change of the orbital semi-major axis and eccentricity due to 
resonances between dynamic tides and free oscillation modes of close binary components. They are functions 
of the orbital eccentricity $e$ and, through the Fourier coefficients $c_{\ell,m,k}$, are proportional to the 
ratio $(R_1/a)^{\ell-2}$. They obey the properties of asymetry $G_{\ell,-m,-k}^{(2)}(e) = -G_{\ell,m,k}^
{(2)}(e)$ and $G_{\ell,-m,-k}^{(3)}(e) = -G_{\ell,m,k}^{(3)}(e)$, and are different from zero only when 
the corresponding coefficients $c_{\ell,m,k}$ are different from zero (see Sect.~\ref{secpot}).  The 
coefficients $G_{\ell,0,0}^{(2)}(e)$ and $G_{\ell,0,0}^{(3)}(e)$ are furthermore identically zero for all 
orbital eccentricities $e$. Similar coefficients $G_{\ell,m,k}^{(1)}(e)$ exist for the rates of secular change 
of the position of the periastron (see Smeyers et al. 1998, Willems 2000, Willems et al. 2003).

\begin{deluxetable*}{lccccc}
\tablewidth{12.0cm}
\tablecolumns{6}
\tablecaption{Coefficients $G_{\ell,m,-m}^{(2)}$ for binaries with circular orbits. \label{G2lmkcirc}}
\tablehead{
    & \colhead{$m=0$} & \colhead{$m= \pm 1$} & \colhead{$m= \pm 2$} & \colhead{$m= \pm 3$} & 
\colhead{$m= \pm 4$}  
   }
\startdata
$\ell=2$ &  0  &  0  & $\mp 3/2$   &  -  &  -   \\
$\ell=3$ &  0  & $\mp (3/8)(R_1/a)$ &  0  & $\mp (15/8)(R_1/a)$ &  -   \\
$\ell=4$ &  0  &  0  & $\mp (15/24)(R_1/a)^2$ &  0  & $\mp (35/16)(R_1/a)^2$ 
\enddata
\end{deluxetable*}

\begin{figure}
\begin{center}
\resizebox{8.0cm}{!}{\includegraphics{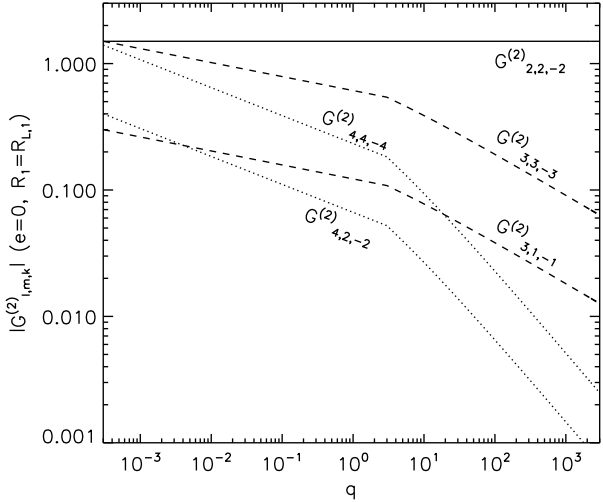}}
\end{center}
\caption{Absolute value of the non-zero coefficients $G_{\ell,m,-m}^{(2)}(e)$ as functions of the mass 
ratio $q = M_2/M_1$, for circular binaries with a Roche-lobe filling primary. Since $G_{\ell,m,-m}^{(2)}
(e) = -G_{\ell,-m,m}^{(2)}(e)$, only the coefficients with positive values of $m$ are shown. }
\label{G2lmkfig2}
\end{figure}

In the particular case of a binary with a circular orbit, the coefficients $G_{\ell,m,k}^{(2)}$ are different 
from zero only when $k=-m$ and $m \ne 0$. The non-zero coefficients take the values
\begin{equation}
G_{\ell,m,-m}^{(2)} = -2\, m\, P_\ell^{|m|}(0)\, c_{\ell,m,-m},
  \label{G2lmke0}
\end{equation}
and are listed in Table~\ref{G2lmkcirc}. If the primary in a binary with a circular orbit furthermore fills its 
Roche lobe, the coefficients $G_{\ell,m,k}^{(2)}$ become functions of the binary mass ratio $q$. The 
variations of these coefficients as a function of the binary mass ratio are shown in Fig.~\ref{G2lmkfig2}. 
For mass ratios $q\gg 1$, the second-degree coefficients $G_{2,m,k}^{(2)}$ are always an order of 
magnitude or more larger than the coefficients $G_{3,m,k}^{(2)}$ and $G_{4,m,k}^{(2)}$, so that the $
\ell=3$ and $\ell=4$ terms can be neglected in the expansion for the rate of secular change of the orbital 
semi-major axis. For mass ratios $q \lesssim 1$, the magnitude of the coefficients $G_{3,m,k}^{(2)}$ and 
$G_{4,m,k}^{(2)}$ can become comparable to the magnitude of the coefficients $G_{2,m,k}^{(2)}$. 
Since non-Roche-lobe filling primaries always have $R_1 < R_{\rm L,1}$, the  curves shown in Fig.~\ref
{G2lmkfig2} pose upper limits on the coefficients $G_{\ell,m,k}^{(2)}$ in detached binaries with circular 
orbits. The coefficients $G_{\ell,m,k}^{(3)}$ are all identically zero for binaries with circular orbits 
(Willems 2000).

In the limiting case where all relevant forcing angular frequencies $\sigma_{m,k}$ are small compared to 
the inverse of the star's dynamical time scale, the quantities $F_{\ell,m,k}$ can be determined by means of 
the quasi-static tide solutions derived in \S\,\ref{qstide}:
\begin{equation}
F_{\ell,m,k} = -{1 \over 2}\, \left[ {R_1 \over {G\,M_1}}\, 
  {{\Psi_{\ell,m,k}^{(0)}(R_1) + \sigma_{m,k}\, \Psi_{\ell,m,k}^{(1)}(R_1)} 
  \over {\varepsilon_{\rm tide}\, c_{\ell,m,k}}} + 1 \right],  \label{Flmkqs}
\end{equation}
where $\Psi_{\ell,m,k}^{(0)}(R_1)$ is real and $\Psi_{\ell,m,k}^{(1)}(R_1)$ is imaginary. The phase 
angles $\gamma_{\ell,m,k}$ are then determined by
\begin{equation}
\tan \gamma_{\ell,m,k} = {{\sigma_{m,k}\, \Psi_{\ell,m,k}^{(1)}(R_1)} 
  \over {\varepsilon_{\rm tide}\, c_{\ell,m,k}}} 
  \left[{{\Psi_{\ell,m,k}^{(0)}(R_1)} 
  \over {\varepsilon_{\rm tide}\, c_{\ell,m,k}}} + \frac{G\, M_1}{R_1} \right]^{-1}. \label{tang}
\end{equation}
When dissipative effects are small, $\tan \gamma_{\ell,m,k} \approx \gamma_{\ell,m,k}$, so that the phase 
angles $\gamma_{\ell,m,k}$ are proportional to the forcing angular frequencies $\sigma_{m,k}$. This 
proportionality is usually assumed in the context of the {\em weak friction approximation} in tidal evolution 
theory (see, e.g., Alexander 1973, Hut 1981, Ruymaekers 1992). In Appendix~\ref{alt}, we show that in the
limiting case of small forcing frequencies and weak damping, Eqs.~(\ref{da4}) and~(\ref{de4}) are 
equivalent to the equations derived by, e.g., Zahn (1977, 1978), Hut (1981), and Ruymaekers (1992).

\section{Stellar Spin Evolution}

Due to 
the phase shift between the perturbation of the gravitational potential and the position of the companion
induced by tidal energy dissipation, the companion exerts a torque $\vec{\cal T}$ on the tidally 
distorted star. The torque is determined from Newton's law of action and reaction as the opposite of the torque 
exerted by the tidally distorted star on the companion due to the perturbation of the star's external 
gravitational potential:
\begin{equation}
\vec{\cal T} = M_2\, (\vec{r} \times \vec{\nabla} \Phi_e^\prime),
\end{equation} 
where $ \Phi_e^\prime$ is to be evaluated at the position of the companion. The tidal torque is perpendicular to the 
orbital plane and has a magnitude 
\begin{eqnarray}
\lefteqn{ {\cal T} = \frac{8\, \pi}{P_{\rm orb}}\, 
   \left( \frac{G\, M_1^2\, M_2^2}{M_1+M_2} \right)^{1/2} 
   \frac{M_2}{M_1}\, a^{1/2} \sum_{\ell=2}^4 
   \sum_{m=-\ell}^\ell \sum_{k=0}^\infty \left( \frac{R_1}{a} \right)^{\ell+3} 
   }  \nonumber \\
 & & \phantom{ww} \times m\, P_\ell^{|m|}(0) \kappa_{\ell,m,k}\, 
   c_{\ell,m,k}\, |F_{\ell,m,k}|  \nonumber \\
 & & \phantom{ww} \times \left( \frac{u}{a} \right)^{-(\ell+1)} 
   \sin \left( m\, v + k\, M  + \gamma_{\ell,m,k} \right).  \hspace{2.5cm}
\end{eqnarray}

\begin{figure}
\begin{center}
\resizebox{8.0cm}{!}{\includegraphics{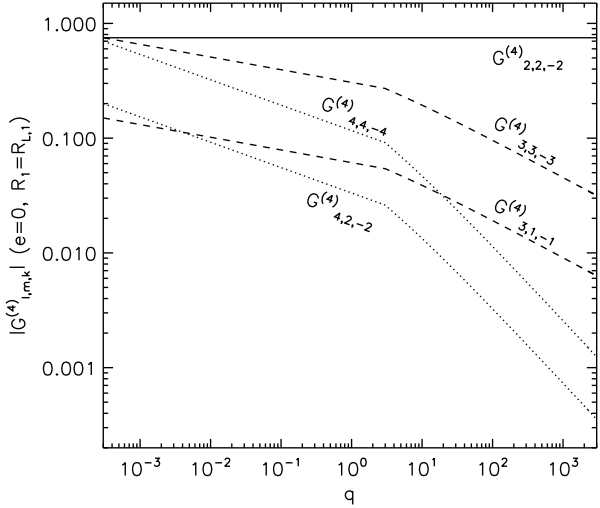}}
\end{center}
\caption{Absolute value of the non-zero coefficients $G_{\ell,m,-m}^{(4)}(e)$ as functions of the mass 
ratio $q = M_2/M_1$, for circular binaries with a Roche-lobe filling primary. Since $G_{\ell,m,-m}^{(4)}
(e) = -G_{\ell,-m,m}^{(4)}(e)$, only the coefficients with positive values of $m$ are shown. }
\label{G4lmkfig2}
\end{figure}

Since we are interested in the long-term secular effects of the tidal torque on the tidally distorted 
star, we average the torque over one revolution of the companion. It follows that 
\begin{eqnarray}
\lefteqn{ \left< {\cal T} \right> =  \frac{8\, \pi}{P_{\rm orb}}\, 
   \left( \frac{G\, M_1^2\, M_2^2}{M_1+M_2} \right)^{1/2} 
   \frac{M_2}{M_1}\, a^{1/2} \sum_{\ell=2}^4 
   \sum_{m=-\ell}^\ell \sum_{k=0}^\infty \left( \frac{R_1}{a} \right)^{\ell+3} 
   }  \nonumber \\
 & & \phantom{ww} \times \kappa_{\ell,m,k}\, |F_{\ell,m,k}|\, 
   \sin \gamma_{\ell,m,k}\, G^{(4)}_{\ell,m,k}(e)   \hspace{3cm}
\end{eqnarray}
with
\begin{equation}
G^{(4)}_{\ell,m,k}(e) = m\, \frac{(\ell + |m|)!}{(\ell - |m|)!}\, 
  \left( \frac{R_1}{a} \right)^{-(\ell-2)} c_{\ell,m,k}^2.
\end{equation}
The coefficients $G^{(4)}_{\ell,m,k}(e)$ are different from zero only for nonaxisymmetric ($m \ne 0$) 
tides and have the same sign as the azimuthal number $m$. They are related to the coefficients $G_
{\ell,m,k}^{(2)}(e)$ and $G_{\ell,m,k}^{(3)}(e)$ as 
\begin{equation}
G^{(4)}_{\ell,m,k}(e) = \frac{e}{(1-e^2)^{1/2}} \left[ G_{\ell,m,k}^{(3)}(e) 
   - \frac{1-e^2}{2\,e}\, G_{\ell,m,k}^{(2)}(e) \right],
\end{equation}
so that their properties can be derived either from the properties of the coefficients $c_{\ell,m,k}$ or from 
the properties of the coefficients $G_{\ell,m,k}^{(2)}(e)$ and $G_{\ell,m,k}^{(3)}(e)$. The values of the 
coefficients $G^{(4)}_{\ell,m,k}$ for binaries with circular orbits are listed in Table~\ref{G4lmkcirc}. The 
variations of the coefficients $G^{(4)}_{\ell,m,k}$ for binaries with circular orbits and Roche-lobe filling 
primaries are shown in Fig.~\ref{G4lmkfig2} as functions of the binary mass ratio. 

\begin{deluxetable*}{lccccc}
\tablewidth{12.0cm}
\tablecolumns{6}
\tablecaption{Coefficients $G_{\ell,m,-m}^{(4)}$ for binaries with circular orbits. \label{G4lmkcirc}}
\tablehead{
    & \colhead{$m=0$} & \colhead{$m= \pm 1$} & \colhead{$m= \pm 2$} & \colhead{$m= \pm 3$} & 
\colhead{$m= \pm 4$}  
   }
\startdata
$\ell=2$ &  0  &  0  & $\pm 3/4$   &  -  &  -   \\
$\ell=3$ &  0  & $\pm (3/16)(R_1/a)$ &  0  & $\pm (15/16)(R_1/a)$ &  -   \\
$\ell=4$ &  0  &  0  & $\pm (5/16)(R_1/a)^2$ &  0  & $\pm (35/32)(R_1/a)^2$ 
\enddata
\end{deluxetable*}

The tidal torque $\vec{\cal T}$ exerted by the companion affects the rotational angular velocity of the 
tidally distorted star. Assuming the star rotates as a rigid body and neglecting any perturbations of the star's 
moment of inertia due to the tidal distortion, the rate of change of the rotational angular velocity $\Omega_1
$ is related to the tidal torque as
\begin{equation}
I_1\, \frac{d\Omega_1}{dt} = {\cal T},
\end{equation}
where $I_1$ is the star's moment of inertia with respect to its rotation axis. Consequently, the rate of secular 
change of the rotational angular velocity $\Omega_1$ is given by
\begin{eqnarray}
\lefteqn{ \left( \frac{d\Omega_1}{dt} \right)_{\rm sec} 
   =  \frac{8\, \pi}{P_{\rm orb}}\, 
   \left( \frac{G\, M_1^2\, M_2^2}{M_1+M_2} \right)^{1/2} 
   \frac{M_2}{M_1}\, \frac{a^{1/2}}{I_1} \sum_{\ell=2}^4 
   \sum_{m=-\ell}^\ell \sum_{k=0}^\infty 
   }  \nonumber \\
 & & \phantom{ww} \times \left( \frac{R_1}{a} \right)^{\ell+3} 
   \kappa_{\ell,m,k}\, |F_{\ell,m,k}|\, 
   \sin \gamma_{\ell,m,k}\, G^{(4)}_{\ell,m,k}(e).  \label{domsec}  \hspace{1.5cm}
\end{eqnarray}

\section{Astrophysical relevance}
\label{appl}

We apply the formalism presented in the previous sections to a $0.3\,M_\odot$ helium white dwarf model 
representative of an isolated white dwarf with a radius of $0.018\,R_\odot$ and an effective temperature of 
3590\,K. Other relevant model properties are summarized in Table~\ref{wdmodels} and details on the model input physics can be found in Deloye et al. (2007). We particularly note that the model has a thin convection zone 
near the stellar surface characterized by a frequency-dependent  turbulent viscosity coefficient 
\begin{equation}
\nu_{m,k} = {L^2 \over \tau_{\rm conv}}\, \left[ 1 + \left( \tau_{\rm conv}\, {\sigma_{m,k} \over {2\, \pi}} 
\right)^s \right]^{-1}.  \label{nu}
\end{equation}
Here, $L$ is the mixing length, $\tau_{\rm conv} = |N^2|^{-1/2}$ the convective turnover time scale
(Terquem et al. 1998, Savonije \& Witte 2002), and $s$ an integer constant. The mixing length is assumed 
to be twice the local pressure scale height. 

\begin{deluxetable}{ll}
\tablecolumns{2}
\tablecaption{White dwarf model properties.  \label{wdmodels}}
\tablehead{ \colhead{Quantity} & \colhead{Value} }
\startdata
$M_1$ & $0.3\,M_\odot$  \\ 
$R_1$ & $0.018\,R_\odot$ \\
$T_{{\rm eff},1}$\tablenotemark{a} & 3590\,K \\
$L_1$ & $4.7 \times 10^{-5}\,L_\odot$ \\ 
$\tau_{{\rm dyn},1}$\tablenotemark{b} & 6.9\,s \\
$\tau_{{\rm HK},1}$\tablenotemark{c} & $3.4 \times 10^{12}$\,yr \\
$\tau_{{\rm dyn},1}/\tau_{{\rm HK},1}$ & $6.5 \times 10^{-20}$ \\
Chemical Composition & helium 
\enddata
\tablenotetext{a}{Effective temperature}
\tablenotetext{b}{Dynamic time scale}
\tablenotetext{c}{Helmholtz-Kelvin time scale}
\end{deluxetable}

The factor between square brackets in Eq.~(\ref{nu}) is a reduction factor to account for the decreased 
efficiency of convective damping when the tidal period $P_{m,k} = 2\,\pi/\sigma_{m,k}$ is shorter than the 
convective turnover time scale $\tau_{\rm conv}$. Because of the dependence of the reduction factor on the 
tidal forcing frequency $\sigma_{m,k}$, the turbulent viscosity coefficient is different for tides generated
by different terms in Expansion~(\ref{W}) for the tide-generating potential (see also Zahn 2008). 
Zahn (1966) proposed a reduction of the turbulent viscosity coefficient characterized by $s=1$, while 
Goldreich \& Keeley (1977) argued for $s=2$ (see also Goldman \& Mazeh 1991, Goodman \& Oh 1997). 
More recently, Penev et al. (2007) calculated the effective turbulent viscosity as a function of the tidal 
forcing frequency using 3-D numerical simulations and found a reduction factor that closely matched the 
$s=1$ prescription proposed by Zahn (1966). In our calculations, we therefore adopt $s=1$. Test 
calculations with $s=2$ show that the choice of $s$ affects our results by less than an order of magnitude. 

\begin{figure}
\begin{center}
\resizebox{8.0cm}{!}{\includegraphics{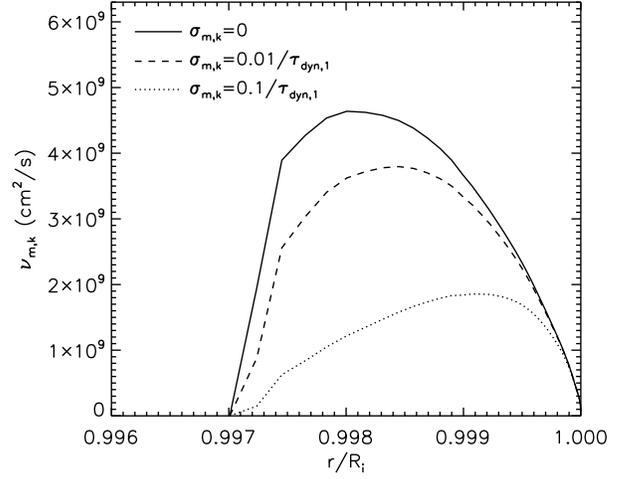}}
\end{center}
\caption{Variation of the turbulent viscosity coefficient $\nu_{m,k}$ in the surface convection 
zone of a $0.3\,M_\odot$ helium white dwarf model for different forcing angular frequencies 
$\sigma_{m,k}$. The $\sigma_{m,k}=0$ line represents the maximum attainable turbulent
viscosity. The dashed and dotted lines represent the reduced turbulent viscosity coefficient for
$\sigma_{m,k}=0.01\, \tau_{{\rm dyn},1}^{-1}$ and $\sigma_{m,k}=0.1\, \tau_{{\rm dyn},1}^{-1}$, 
assuming $s=1$.}
\label{nufig}
\end{figure}

In Fig.~\ref{nufig}, we show the variations of the turbulent viscosity coefficient $\nu_{m,k}$ in the 
$0.3\,M_\odot$ helium white dwarf model as a function of the normalized radial coordinate $r/R_1$. The 
maximum attainable turbulent viscosity is indicated by the $\sigma_{m,k}=0$ line. The dashed and dotted 
lines show the reduced  turbulent viscosity for $s=1$ and tidal forcing frequencies $\sigma_{m,k} = 
0.01\, \tau_{{\rm dyn},1}^{-1}$ and $\sigma_{m,k} = 0.1\, \tau_{{\rm dyn},1}^{-1}$, respectively. 
The reduction of the turbulent viscosity coefficient is most prominent near the base of the convection 
zone where the convective turnover time scale is longest. 

\begin{figure}
\begin{center}
\resizebox{8.0cm}{!}{\includegraphics{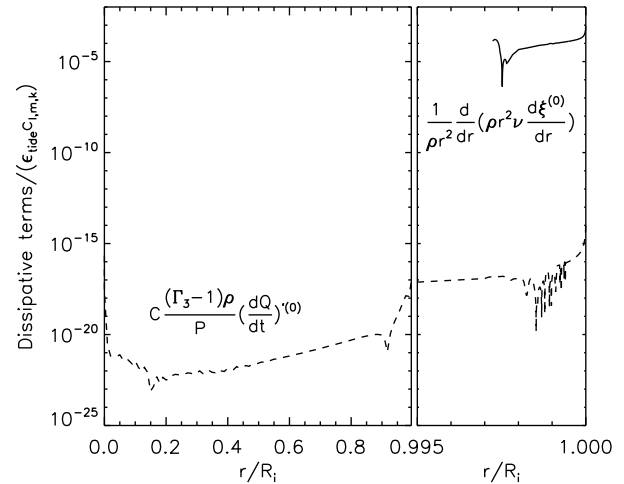}}
\end{center}
\caption{Order of magnitude of the convective and radiative damping terms in the system of differential 
equations governing quasi-static tides in a $0.3\,M_\odot$ helium white dwarf model. The terms are 
calculated for $\ell=2$ and are rendered dimensionless by expressing the physical quantities in the 
units listed in table~\ref{units}. The convective damping term is calculated for a forcing angular frequency 
$\sigma_{m,k}=0$. }
\label{nafig}
\end{figure}

For non-degenerate stars dissipation of tidal energy through quasi-static tides is dominated by 
convective damping (Zahn 1977). This still holds true in white dwarfs, as illustrated in Fig.~\ref{nafig} 
where the variations of the radiative and convective damping terms
\[
\frac{1}{\rho\, r^2}\, \frac{d}{dr} \left( \rho\, r^2\, \nu_{m,k} \frac{d\xi^{(0)}}{dr} \right)
\phantom{ww} \mbox{and} \phantom{ww} 
C\, \frac{(\Gamma_3-1) \rho}{P} \left( \frac{dQ}{dt} \right)^{\! \prime (0)}
\]
are shown as functions of the normalized radial coordinate $r/R_1$. The terms are obtained 
by numerically integrating differential Eq.~(\ref{psi0}) for $\ell=2$, and using Eqs.~(\ref{zero}) 
and~(\ref{dQdt0}) to determine $\xi^{(o)}(r)$ and $(dQ/dt)^{(o)}$ from the lowest-order total perturbation of the 
gravitational potential $\Psi^{(o)}(r)$.  For the determination of the turbulent viscosity term $\nu_{m,k}$, the forcing
angular frequency $\sigma_{m,k}$ was set equal to zero. The terms are furthermore rendered dimensionless
by expressing the physical quantities in the units listed in Table~\ref{units}. For convenience, we also 
divided the terms by the scaling factor $\varepsilon_{\rm tide}\, c_{\ell,m,k}$, so that the curves shown in the 
figure are independent of the binary orbital period, eccentricity, and companion mass. In the convection zone, the 
turbulent damping term exceeds the radiative damping term by more than 10 orders of magnitude. We could 
therefore safely have set the constant $C^\prime$ equal to 
zero in the derivation of the quasi-static tide solutions presented in \S\,\ref{qstide}.

In the following subsections we calculate the tidal distortion and the orbital evolution time scales due to 
quasi-static tides for both detached and semi-detached white dwarf binaries
and for different ranges of binary component masses, orbital periods, orbital eccentricities, and white 
dwarf rotation rates. We only consider mass ratios $q \ge 1$, so that $\ell=3$ and $\ell=4$ terms can
be neglected in the expansion of the tide-generating potential (see \S\,\ref{secpot}).

\subsection{Detached binaries}

We first turn our attention to tidal interactions in detached white dwarf binaries. Given the strong 
dependence of tidal effects on the ratio of the white dwarf radius to the orbital semi-major axis, we 
focus on short-period binaries with orbital frequencies relevant to the Laser Interferometer
Space Antenna, LISA ($10^{-4}$--$10^{-1}$\,Hz). In all cases, the companion star is assumed to be 
more compact than the white dwarf so that we do not have to worry about Roche-lobe overflow from the 
companion. As indicated above, we also restrict ourselves to the dominant quadrupole tides generated by 
the $\ell=2$ terms in Expansion~(\ref{W}) of the tide-generating potential. 

To calculate the tidal distortion of the $0.3\,M_\odot$ helium white dwarf model, we solve Eqs.~(\ref{psi0}) 
and~(\ref{psi1}) for $\Psi^{(0)}(r)$ and $\Psi^{(1)}(r)$ with their respective boundary conditions for each 
non-zero and non-negligible term in Expansion~(\ref{W}) of the tide-generating potential. Equations~(\ref{psi0}) 
and~(\ref{psi1}) are solved using a variable step Runge-Kutta integrator, following the 
semi-analytical procedure outlined in Appendix~\ref{solution}. The radial component of the tidal 
displacement field and the Eulerian perturbations of the stellar structure quantities are then determined 
from Eqs.~(\ref{zero}), (\ref{f1})--(\ref{f2}), and (\ref{Tp0})--(\ref{Tp1}). We recall that the inclusion of 
dissipative effects renders the solutions complex and introduces a phase shift between the tidal 
perturbations and the tide-generating potential. The amplitude of the tidal perturbations is dominated 
by the real part of the solutions. 

The total tidal displacement field and perturbations of the stellar structure quantities are obtained by 
adding the solutions associated with all non-zero and non-negligible terms in the expansion of the 
tide-generating potential. For instance, the total perturbation of the gravitational potential due to 
quasi-static tides is given by   
\begin{eqnarray}
\lefteqn{\Psi_{\rm tide} \left( \vec{r}, t \right) = \displaystyle 2 \sum_{\ell=2}^4
  \sum_{m=-\ell}^\ell \sum_{k=0}^\infty \kappa_{\ell,m,k} \left| \Psi_{\ell,m,k}(r) 
  \right| P_\ell^{|m|} (\cos \theta) } \nonumber \\
 & & \phantom{ww} \times 
  \cos \left[ m\, \phi + \sigma_{m,k}\, t - k\, n\, \tau + \Upsilon_{\Psi_{\ell,m,k}}(r) \right], 
  \hspace{1.5cm}
\end{eqnarray}
where
\begin{equation}
\Upsilon_{\Psi_{\ell,m,k}}(r) = \arctan \frac{\Im[\Psi_{\ell,m,k}(r)]}{\Re[\Psi_{\ell,m,k}(r)]}.
  \hspace{1.5cm}
\end{equation}
The total radial component of the tidal displacement field and the total Eulerian perturbations of the stellar 
structure quantities are obtained similarly.

\begin{figure}
\begin{center}
\resizebox{8.0cm}{!}{\includegraphics{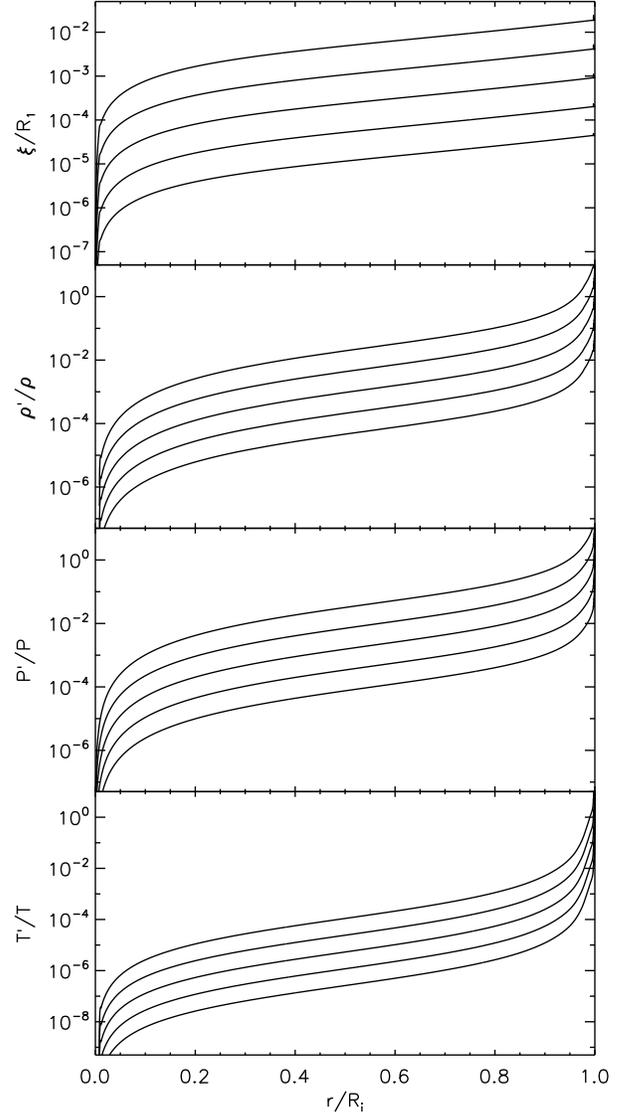}}
\end{center}
\caption{Amplitude of the radial component $\xi$ of the tidal displacement field and the Eulerian 
perturbations of the mass density $\rho$, pressure $P$, and temperature $T$ for quasi-static tides 
associated with $\ell=2$ in a $0.3\,M_\odot$ helium white dwarf model. The white dwarf is assumed 
to have a rotation period of 100\,hr and to have a $0.3\,M_\odot$ point-mass companion in a circular orbit. 
From top to bottom, the different curves in each panel correspond to orbital periods $P_{\rm orb} = 3$, 
6, 13, 28, and 60\, min.}
\label{tidfig}
\end{figure}

In Fig.~\ref{tidfig}, the amplitudes of the radial component of the tidal displacement field and the 
Eulerian perturbations of the mass density, pressure, and temperature are shown as functions of the
normalized radial coordinate $r/R_1$. The $0.3\,M_\odot$ helium white dwarf is assumed to have an equal mass  
companion ($q=1$) in a circular binary with orbital periods ranging from 3 to 60 minutes. The white dwarf
rotation period is assumed to be 100 hours. As expected, the amplitudes of the perturbations increase 
outward and are larger for shorter orbital periods. The radial component of the tidal displacement field 
at the star's  surface in particular decreases from $2 \times 10^{-2}\, R_1$ to $4 \times 10^{-5}\, R_1$ 
when the orbital period increases from 3 to 60\,min.

\begin{figure}
\begin{center}
\resizebox{8.0cm}{!}{\includegraphics{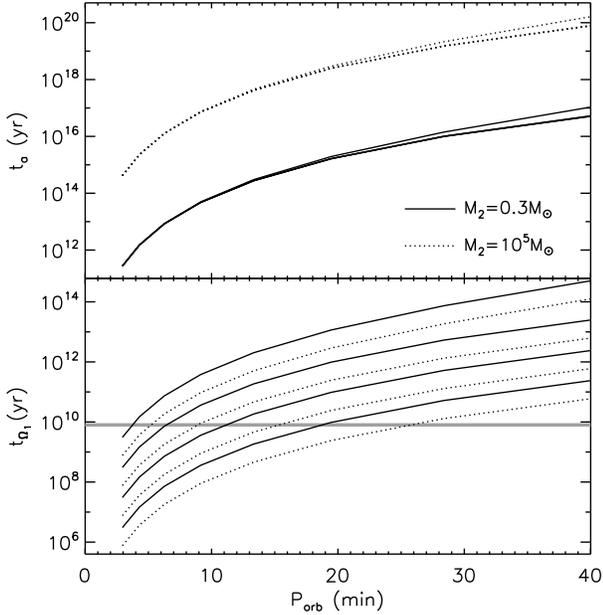}}
\end{center}
\caption{Tidal evolution time scales for the rates of secular change of the orbital semi-major axis $a$ and 
the white dwarf rotational angular velocity $\Omega_1$ for a $0.3\,M_\odot$ helium white dwarf in a 
circular binary. Solid and dotted lines represent time scales for point-mass companions of $0.3\,M_\odot$
and $10^5\,M_\odot$, respectively. From top to bottom, the solid and dotted lines in each panel correspond
to white dwarf rotation periods of 1, 10, 100, and 1000\,hr. In the top panel, the curves associated with 
rotation periods longer than 10\,hr are indistinguishable. The grey horizontal line represents the age of
an isolated $0.3\,M_\odot$ helium white dwarf when it has cooled to a temperature of 3590\,K. }
\label{orbfig}
\end{figure}

Next, we consider the tidal evolution time scales $t_a = |\dot{a}/a|$ and $t_{\Omega_1} = 
|\dot{\Omega}_1/\Omega_1|$ for the orbital semi-major axis and the white dwarf rotational angular 
velocity. The time scales are shown in Fig.~\ref{orbfig} as a function of the orbital period for a
$0.3\,M_\odot$ helium white dwarf in a circular orbit around a  $0.3\,M_\odot$ or $10^5\,M_\odot$ 
point-mass companion\footnote{We note that at the shortest orbital periods shown ($P_{\rm orb} 
\approx 3$\,min), the forcing angular frequencies $\sigma_{m,k}$ can be close to $0.5\, \tau_{{\rm dyn},1}$, 
stretching the applicability of the applied perturbation theory.}. The different curves correspond to 
different white dwarf rotation periods ranging from 1 to 1000\,hr.

\begin{figure}
\begin{center}
\resizebox{8.0cm}{!}{\includegraphics{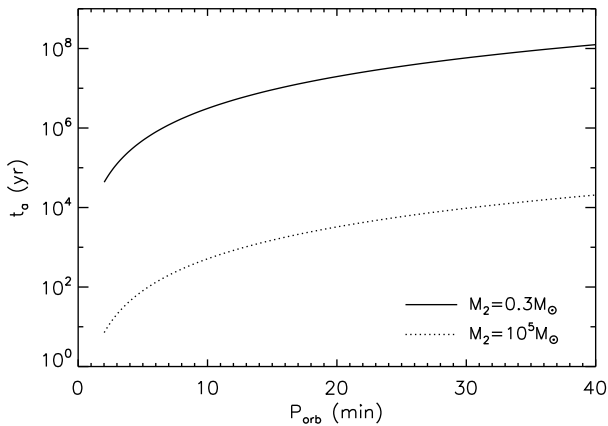}}
\end{center}
\caption{Time scales for the rate of secular change of the orbital semi-major axis $a$ due to gravitational
wave emission for a $0.3\,M_\odot$ white dwarf with a $0.3\,M_\odot$ or $10^5\,M_\odot$ companion in 
a circular orbit.}
\label{grfig}
\end{figure}

The time scales for the rate of secular change of the orbital semi-major axis are longer than a Hubble 
time for all binary configurations considered. They increase with increasing binary companion mass 
due to the associated increase of the orbital semi-major axis for a given orbital 
period. For comparison, the time scales of orbital evolution due to gravitational wave emission determined 
from the Peters (1964) equations are shown in Fig.~\ref{grfig} for the same binary component masses 
and orbital period range as used in Fig.~\ref{orbfig}. The time scales of orbital evolution due to 
convective damping of quasi-static tides are several orders of magnitude larger than those due to 
gravitational radiation, so that the effects of quasi-static tides in white dwarfs can safely be neglected 
when predicting gravitational wave signals from white dwarf binaries. 

The time scales for the rate of secular change of the white dwarf rotational angular velocity are at least two 
orders of magnitude shorter than the time scales for the rates of secular change of the orbital semi-major 
axis. For a given orbital period, they decrease significantly with increasing white dwarf rotation period or, 
equivalently, with increasing degree of asynchronism. For white dwarf rotation periods of 1000\,hr, the time 
scales become shorter than the age of an isolated $0.3\,M_\odot$ helium white dwarf of 3590\,K for orbital 
periods below $\sim 20$--25\,min, depending on the white dwarf companion mass. However, for all binary 
configurations considered, the white dwarf spin-up time scales are still considerably longer than the orbital 
evolution time scales due to gravitational radiation. Quasi-static tides will therefore not be able to spin the 
white dwarf up fast enough to reach a synchronous rotation rate.

\begin{figure}
\begin{center}
\resizebox{8.0cm}{!}{\includegraphics{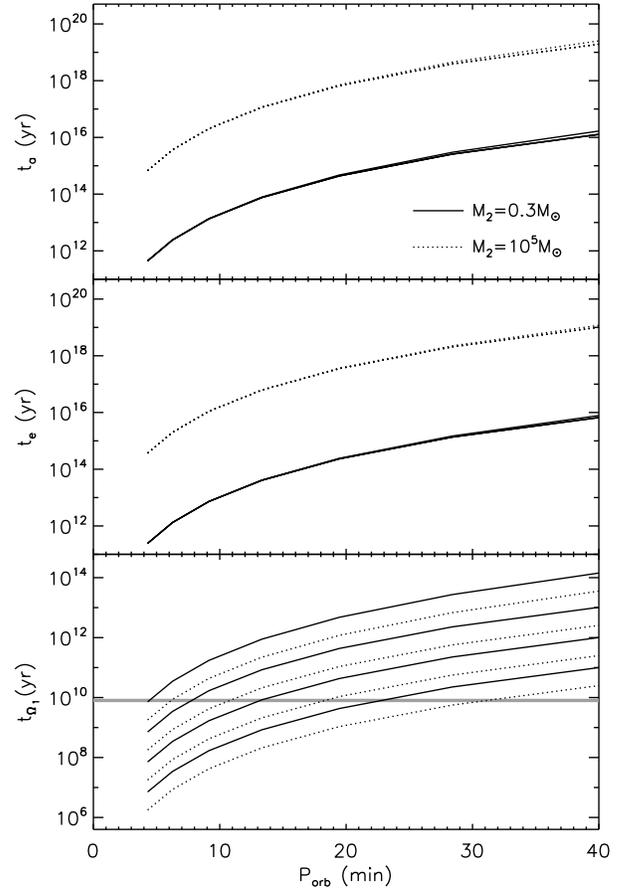}}
\end{center}
\caption{Tidal evolution time scales for the rates of secular change of the orbital semi-major axis $a$,
the orbital eccentricity $e$, and the white dwarf rotational angular velocity $\Omega_1$ for a $0.3\,M_\odot$ 
helium white dwarf in a binary with orbital eccentricity $e=0.3$. The different lines have the same meaning 
as in Fig.~\ref{orbfig}.  In the top two panels, the lines associated with rotation periods longer than 10\,hr 
are indistinguishable. }
\label{orbeccfig}
\end{figure}

The effects of the orbital eccentricity on the tidal evolution time scales are illustrated in Fig.~\ref{orbeccfig}, 
where the tidal evolution time scales $t_a = |\dot{a}/a|$, $t_e = |\dot{e}/e|$, and 
$t_{\Omega_1} = |\dot{\Omega}_1/\Omega_1|$ are shown for a $0.3\,M_\odot$ helium white dwarf orbiting 
a $0.3\,M_\odot$ or $10^5\,M_\odot$ point-mass companion with an orbital eccentricity $e=0.3$. 
For a given orbital period, the time scales are shorter than those for a circular binary due to the stronger tidal 
interactions taking place at the periastron of the binary orbit. However, the time scales for the rates of secular
change of the orbital semi-major axis and eccentricity remain longer than a Hubble time for all binary
configurations considered. Depending on the binary companion mass, the time scales for the rate of  
secular change of the white dwarf's rotational angular velocity can become shorter than the age of an 
isolated $0.3\,M_\odot$ helium white dwarf of 3590\,K for orbital periods $P_{\rm orb} \la 30$\,min, 
provided the initial degree of asynchronism at periastron is sufficiently high.

\subsection{Mass-transferring binaries}

Next, we consider tidal interactions for binaries in which the considered $0.3\,M_\odot$ helium 
white dwarf fills its Roche lobe. We ignore any coupling between mass transfer and tides and 
assume the radius of the white dwarf to be exactly equal to the radius of its Roche lobe. The 
orbital separation is then fully determined by the mass ratio by means of Eq.~(\ref{RL}) for
the volume-equivalent radius of the white dwarf's Roche lobe. We refrain from comparing the
tidal evolution time scales with the time scales of orbital evolution due to mass transfer since 
the determination of the latter requires detailed tracking of the response of the white dwarf to 
mass loss (see, e.g., Deloye \& Taam 2006, Deloye et al. 2007).

\begin{figure}
\begin{center}
\resizebox{8.0cm}{!}{\includegraphics{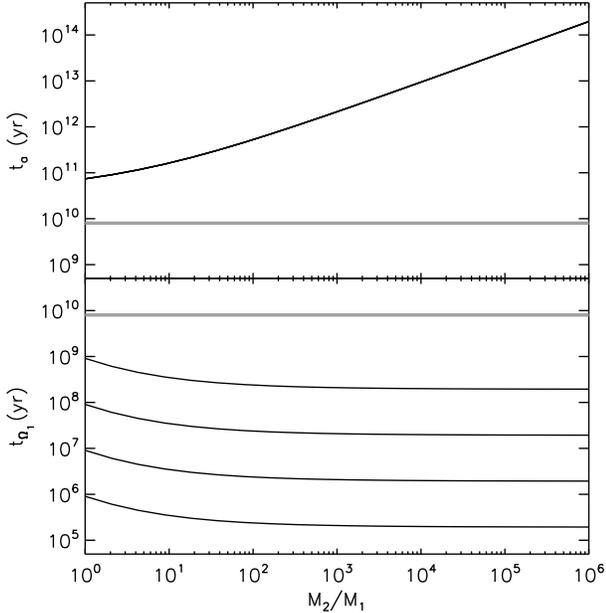}}
\end{center}
\caption{Tidal evolution time scales for the rates of secular change of the orbital semi-major axis $a$ 
and the white dwarf rotational angular velocity $\Omega_1$ as a function of the binary mass ratio 
$q=M_2/M_1$, for a Roche-lobe filling $0.3\,M_\odot$ helium white dwarf in a circular binary. From 
top to bottom, the lines in the bottom panel correspond to white dwarf rotation periods of 1, 10, 100, 
and 1000\,hr. In the top panel, lines associated with the rotation periods are indistinguishable. The grey
horizontal line represents the age of an isolated $0.3\,M_\odot$ helium white dwarf when it has cooled 
to a temperature of 3590\,K. }
\label{orbMTfig}
\end{figure}

The time scales $t_a = |\dot{a}/a|$ and $t_{\Omega_1} = |\dot{\Omega}_1/\Omega_1|$ for the rates
of secular change of the orbital semi-major axis and the white dwarf rotational angular velocity
are shown in Fig.~\ref{orbMTfig} as functions of the binary mass ratio, for a Roche-lobe
filling $0.3\,M_\odot$ helium white dwarf in a circular binary. The different curves in the bottom panel 
of the figure correspond to white dwarf rotation periods ranging from 1 to 1000\,hr. The tidal evolution 
time scales for the orbital semi-major axis are orders of magnitude larger than those due to gravitational 
radiation, so that the effects of quasi-static tides on gravitational wave signals from white dwarf binaries can be 
neglected even for semi-detached white dwarf binary systems such as AM CVn binaries. The tidal 
spin-up time scale, on the other hand, can be shorter than 1\,Myr, provided that the white dwarf is 
rotating much slower than the orbital motion of the companion.

\section{Concluding remarks}

We derived a formalism to study dissipative tides in white dwarfs in the limiting case of 
quasi-static tides. The limit is applicable to binaries of arbitrary eccentricities as long as the forcing 
angular frequencies of the non-negligible terms in the Fourier expansion of the tide-generating 
potential are smaller than the inverse of the white dwarf's dynamical time scale. We account for both 
convective and radiative damping of quasi-static tides, but find the total perturbation of the 
gravitational potential, 
which determines the binary's tidal evolution, to be affected mainly by convective damping. At the 
order of approximation considered, radiative damping affects the total perturbation of the gravitational 
potential only through the surface boundary condition expressing the continuity of the gravitational 
potential and its gradient at the star's surface. 

We applied the formalism to binaries consisting of a 
$0.3\,M_\odot$ helium white dwarf and a $0.3\,M_\odot$ or $10^5\,M_\odot$ point-mass companion, 
representative of double degenerates (e.g. Nelemans et al. 2001) and coalescing white 
dwarf--massive black hole binaries (Menou, Haiman, \& Kocsis 2008; Sesana et al. 2009), respectively.  
Since the evolutionary time scales depend strongly on the ratio of the white dwarf radius to the 
orbital semi-major axis, we focused the application on short-period binaries with orbital frequencies in 
LISA gravitational wave frequency band ($10^{-4}$--$10^{-1}$\,Hz). The time scales for the rate  
of secular change of the orbital semi-major axis and eccentricity are found to be longer than a Hubble 
time for all binary configurations considered. Orbital evolution due to convective damping of 
quasi-static tides can therefore be neglected in the construction of gravitational wave templates of white 
dwarf binaries for LISA. The time scales for the rate of secular change of the white dwarf's rotational
angular velocity, on the other hand, can be shorter than 10\,Myr, especially if the white dwarf is initially 
rotating with a frequency that is much smaller than the binary orbital frequency.

Even though tidal spin-up of the white dwarf can occur in astrophysically interesting time spans, 
the time scales are still much longer than the orbital inspiral time scale due to gravitational radiation. 
Tides will therefore not be able to catch up with gravitational-radiation driven orbital evolution and 
synchronize the white dwarf's rotation with the orbital motion. White dwarf binaries in the LISA 
band therefore naturally evolve from a low- to a high-frequency tidal forcing regime. For the shortest period 
binaries considered, the quasi-static tide approximation therefore breaks down and energy dissipation 
through dynamic tides and, in particular, resonantly excited $g$-modes must be taken into account. 
Possible mechanisms contributing to damping of nonradial $g$-modes are radiative heat leakage, 
neutrino losses, and gravitational radiation (Osaki \& Hansen 1973; Rathore, Blandford, \& Broderick 
2005), or, in the case of large amplitude modes, nonlinear coupling to other nonradial oscillation 
modes (e.g.\ Dziembowski 1982; van Hoolst 1994ab, Wu \& Goldreich 2001). We will explore energy 
dissipation through dynamic tides in detail in future investigations.

\acknowledgments This research is supported by NASA-BEFS grant NNG06GH87G and a Packard 
Fellowship in Science and Engineering to Vicky Kalogera. Numerical 
simulations were performed on the HPC cluster {\tt Fugu} available to the Theoretical Astrophysics 
Group at Northwestern University through NSF MRI grant PHY-0619274 to Vicky Kalogera.

\appendix
\section{A. Perturbation of the temperature}
\label{temp}

Assuming the mean molecular weight of a mass element does not change on time scales comparable to or 
shorter than a tidal oscillation period, the Lagrangian perturbation of the entropy $S$ due to a star's tidal 
distortion can be related to the Lagrangian perturbations of the temperature $T$ and mass density $\rho$ as 
(Unno et al. 1989, Eq. 13.74)
\begin{equation}
(\delta S)_{\rm tide} = C_V\, \left[ \frac{(\delta T)_{\rm tide}}{T} 
  - \left( \Gamma_3 - 1 \right) \frac{(\delta \rho)_{\rm tide}}{\rho} \right],
\end{equation}
where $C_V$ is the specific heat per unit mass at constant density, and a $\delta$ in front of a quantity 
denotes the Lagrangian perturbation of that quantity. By taking the total time derivative of this expression, 
using the property that the total time derivative commutes with the Lagrangian perturbation operator, and 
substituting $dS/dt = (1/T)(dQ/dt)$, it follows that
\begin{equation}
\left[ \delta \left( \frac{1}{T}\, \frac{dQ}{dt} \right) \right]_{\rm tide} 
   = C_V\, \left[ \frac{d}{dt} \left( \frac{\delta T}{T} \right)_{\rm tide} 
   - \left( \Gamma_3 - 1 \right) \frac{d}{dt} \left( \frac{\delta \rho}{\rho} \right)_{\rm tide} \right].
\end{equation}

Next, passing on from Lagrangian to Eulerian perturbations and taking into account that the unperturbed star 
is static and in thermal equilibrium yields
\begin{equation}
 \frac{1}{T} \left( \frac{dQ}{dt} \right)^\prime_{\rm tide} = C_V\, \left[ \frac{\partial}{\partial t}
    \left( \frac{T^\prime_{\rm tide}}{T} + \frac{1}{T}\, \frac{dT}{dr}\, \xi_{\rm tide} \right) 
   - \left( \Gamma_3 - 1 \right) \frac{\partial}{\partial t} \left( \frac{\rho^\prime_{\rm tide}}{\rho}
    + \frac{1}{\rho}\, \frac{d\rho}{dr}\, \xi_{\rm tide} \right) \right]. 
\end{equation}
Finally, after separating the time $t$ and angular coordinates $\theta$ and $\phi$ by expanding the tidal 
displacement field and the perturbed stellar structure quantities in Fourier series similar to those used in \S\,
\ref{goveqs}, and eliminating the Eulerian perturbation of the mass density by means of Eq.~(\ref{sep4}), 
the radial part of the Eulerian perturbation of the temperature associated with the spherical harmonic $Y_
\ell^m(\theta,\phi)$ and the forcing angular frequency $\sigma_{m,k}$ is given by
\begin{equation}
\frac{T^\prime_{\ell,m,k}}{T} = - \frac{1}{T}\, \frac{dT}{dr}\, \xi_{\ell,m,k} - \left( \Gamma_3 - 1 
\right) \alpha_{\ell,m,k} 
   - {\rm i}\, \frac{1}{\sigma_{m,k}\, C_V\, T} \left( \frac{dQ}{dt} \right)^\prime_{\ell,m,k}.  \label{Tp}
\end{equation}

In analogy with Eq.~(\ref{sep5dim}), Eq.~(\ref{Tp}) can be cast in dimensionless form by expressing 
the physical quantities in the units listed in Table~\ref{units}. In addition, the temperature $T$ and 
specific heat $C_V$ are expressed in the units $G\,M_1/({\cal R}_{\rm gas}\, R_1)$ and 
${\cal R}_{\rm gas}$, where ${\cal R}_{\rm gas}$ 
is the universal gas constant. It follows that 
\begin{equation}
\frac{T^\prime_{\ell,m,k}}{T} = - \frac{1}{T}\, \frac{dT}{dr}\, \xi_{\ell,m,k} 
   - \left( \Gamma_3 - 1 \right) \alpha_{\ell,m,k} 
   - {\rm i}\, \sigma_{m,k}\, C^\prime\, \frac{1}{C_V\, T} 
   \left( \frac{dQ}{dt} \right)^\prime_{\ell,m,k},
\end{equation}
where $C^\prime$ is defined by Eq.~(\ref{C}). At ${\cal O}(\sigma_{m,k}^0)$, the Eulerian perturbation of the temperature is then given by
\begin{equation}
\frac{T^{\prime (0)}_{\ell,m,k}}{T} = - \frac{1}{T}\, \frac{dT}{dr}\, \xi^{(0)}_{\ell,m,k},   
  \label{Tp0}
\end{equation}
and at ${\cal O}(\sigma_{m,k})$ by
\begin{equation}
\frac{T^{\prime (1)}_{\ell,m,k}}{T} = - \frac{1}{T}\, \frac{dT}{dr}\, \xi^{(1)}_{\ell,m,k}
   - \left( \Gamma_3 - 1 \right) \alpha_{\ell,m,k}^{(1)} 
   - {\rm i}\, C^\prime\, \frac{1}{C_V\, T} 
   \left( \frac{dQ}{dt} \right)^{\! \prime (0)}_{\ell,m,k}.   \label{Tp1}
\end{equation}

\section{B. Perturbation of the rate of change of thermal energy}
\label{thermal}

The application of boundary Condition~(\ref{bcpsi1}) for the total perturbation of the gravitational potential 
and its first derivative requires the evaluation of the Eulerian perturbation of the rate of change of thermal 
energy at the star's surface. Denoting the rate of energy generation per unit mass by $\varepsilon$ and the 
energy flux by $\vec{F}$, the Eulerian perturbation of the rate of change of thermal energy is given by
\begin{equation}
\left( {{dQ} \over {dt}} \right)^{\! \prime}_{\rm tide} = \varepsilon^\prime_{\rm tide} + \frac{\rho^
\prime_{\rm tide}}{\rho^2}\, 
   \vec{\nabla} \cdot \vec{F} - \frac{1}{\rho}\, \vec{\nabla} \cdot \vec{F}^\prime_{\rm tide},
   \label{ddQ}
\end{equation}
where
\begin{equation}
\vec{F} = - \frac{4\, a\, c}{3}\, \frac{T^3}{\kappa\, \rho}\, \vec{\nabla} T.  \label{flux}
\end{equation}
In the latter equation, $a$ is the radiation constant, $c$ the speed of light, $T$ the temperature, and $\kappa
$ the opacity. By using Eq.~(\ref{flux}), we implicitly assume the energy transfer in the near the star's 
surface to be radiative and/or conductive (e.g.\ Hansen, Kawaler, \& Trimble 2004). Since a proper 
treatment of the contribution of convection to the perturbation of the rate of change of thermal energy 
requires a non-local 
time-dependent theory of convection, which is beyond the scope of this investigation, we here simply 
neglect convective effects in Eq.~(\ref{ddQ}) for $(dQ/dt)_{\rm tide}^\prime$. In the following 
derivation, all quantities are assumed to be expressed in the same units as those adopted in \S\,\ref{qstide} 
(see Table~\ref{units}). 

The Eulerian perturbation of the energy generation rate can be expressed in terms of the Eulerian 
perturbations of the mass density $\rho$, the temperature $T$, and the mean molecular weight $\mu$ as
\begin{equation}
\frac{\varepsilon^\prime_{\rm tide}}{\varepsilon} = \varepsilon_\rho\, \frac{\rho^\prime_{\rm tide}}{\rho} 
   + \varepsilon_T\, \frac{T^\prime_{\rm tide}}{T} 
   + \varepsilon_\mu\, \frac{\mu^\prime_{\rm tide}}{\mu}, 
\end{equation}
where
\begin{equation}
\varepsilon_\rho = \left( \frac{\partial \ln \varepsilon}{\partial \ln \rho} \right)_{T,\mu},  \phantom{www}
\varepsilon_T =  \left( \frac{\partial \ln \varepsilon}{\partial \ln T} \right)_{\rho,\mu},   \phantom{www}
\varepsilon_\mu =  \left( \frac{\partial \ln \varepsilon}{\partial \ln \mu} \right)_{\rho,T}.
\end{equation}
If diffuse mixing is assumed to be negligible on time scales of the order of the tidal oscillation period or 
shorter, the Lagrangian perturbation of the mean molecular weight is zero, so that
\begin{equation}
\mu^\prime_{\rm tide} = - \frac{d\mu}{dr}\, \xi_{\rm tide}  \label{mup}
\end{equation}
(e.g.\ Savonije \& Papaloizou 1984, Witte \& savonije 1999). 

From Eq.~(\ref{flux}) and the spherical symmetry of the unperturbed star, it follows that the Eulerian 
perturbation of the energy flux is given by
\begin{equation}
\vec{F}^\prime_{\rm tide} = F \left( \frac{dT}{dr} \right)^{-1}\, \left[ \left( 3\, \frac{T^\prime_{\rm 
tide}}{T}
   - \frac{\kappa^\prime_{\rm tide}}{\kappa} - \frac{\rho^\prime_{\rm tide}}{\rho} \right) \vec{\nabla} T 
   + \vec{\nabla} T^\prime_{\rm tide} \right]. 
\end{equation}
In this equation, the Eulerian perturbation of the opacity can be expressed in terms of the Eulerian 
perturbations of the mass density $\rho$, the temperature $T$, and the mean molecular weight $\mu$ as
\begin{equation}
\frac{\kappa^\prime_{\rm tide}}{\kappa} = \kappa_\rho\, \frac{\rho^\prime_{\rm tide}}{\rho} 
   + \kappa_T\, \frac{T^\prime_{\rm tide}}{T} 
   + \kappa_\mu\, \frac{\mu^\prime_{\rm tide}}{\mu},   \label{kapp}
\end{equation}
where
\begin{equation}
\kappa_\rho = \left( \frac{\partial \ln \kappa}{\partial \ln \rho} \right)_{T,\mu},  \phantom{www}
\kappa_T =  \left( \frac{\partial \ln \kappa}{\partial \ln T} \right)_{\rho,\mu},   \phantom{www}
\kappa_\mu =  \left( \frac{\partial \ln \kappa}{\partial \ln \mu} \right)_{\rho,T}.
\end{equation}

After separating the time $t$ and the angular coordinates $\theta$ and $\phi$ by expanding the tidal 
displacement field and the perturbed stellar structure quantities in Fourier series similar to those used in \S\,
\ref{goveqs}, and using Eqs.~(\ref{mup})--(\ref{kapp}), it follows that
\begin{eqnarray}
\lefteqn{ \vec{\nabla} \cdot \vec{F}^\prime_{\ell,m,k} = \frac{1}{r^2}\, \frac{d}{dr} 
   \left\{ r^2\, F \left[ \left( 3 - \kappa_T \right)\, 
   \frac{T^\prime_{\ell,m,k}}{T} - \left( 1 + \kappa_\rho \right)\,  \frac{\rho^\prime_{\ell,m,k}}{\rho}  
   + \kappa_\mu\, \frac{1}{\mu}\, \frac{d\mu}{dr}\, \xi_{\ell,m,k}   
   + \left( \frac{dT}{dr} \right)^{-1} \frac{dT^\prime_{\ell,m,k} }{dr} \right] \right\}  } \nonumber \\
 & & \phantom{www}  - \frac{\ell (\ell+1)}{r^2}\, F\, \left( \frac{dT}{dr} \right)^{-1} T^\prime_
{\ell,m,k}. 
   \hspace{8cm}
\end{eqnarray}
Consequently, the Eulerian perturbation of the rate of change of thermal energy takes the form
\begin{eqnarray}
\lefteqn{ \left( {{dQ} \over {dt}} \right)^{\! \prime}_{\ell,m,k}  = 
    \varepsilon \left( \varepsilon_\rho\, \frac{\rho^\prime_{\ell,m,k} }{\rho} 
   + \varepsilon_T\, \frac{T^\prime_{\ell,m,k} }{T}  - \varepsilon_\mu\, \frac{1}{\mu}\, \frac{d\mu}{dr}\, 
\xi_{\ell,m,k}  \right)
   + \left[ \frac{1}{\rho\, r^2}\, \frac{d}{dr} \left( r^2\, F \right) \right] \frac{\rho^\prime_{\ell,m,k} }
{\rho} 
   + \frac{1}{\rho}\, \frac{\ell (\ell+1)}{r^2}\, F\, \left( \frac{dT}{dr} \right)^{-1} T^\prime_{\ell,m,k} } 
\nonumber \\
 & & \phantom{www}  - \frac{1}{\rho\, r^2}\, \frac{d}{dr} \left\{ r^2\, F \left[ \left( 3 - \kappa_T \right)\, 
   \frac{T^\prime_{\ell,m,k} }{T} - \left( 1 + \kappa_\rho \right)\,  \frac{\rho^\prime_{\ell,m,k} }{\rho}  
   + \kappa_\mu\, \frac{1}{\mu}\, \frac{d\mu}{dr}\, \xi_{\ell,m,k}   
   + \left( \frac{dT}{dr} \right)^{-1} \frac{dT^\prime_{\ell,m,k} }{dr} \right] \right\}.  \hspace{3cm}
\end{eqnarray}

Next, introducing expansions of the form given by Eq.~(\ref{fexp}), substituting the solutions for $\rho^
{\prime (0)}$ and $T^{\prime (0)}$ given by Eqs.~(\ref{zero}) and~(\ref{Tp0}), and using the chain rule 
yields
\begin{eqnarray}
\lefteqn{ \left( {{dQ} \over {dt}} \right)^{\! \prime (0)}_{\ell,m,k} = - \frac{d \varepsilon}{dr}\, \xi^
{(0)}_{\ell,m,k} 
   - \left[ \frac{1}{\rho\, r^2}\, \frac{d}{dr} \left( r^2\, F \right) \right] \frac{1}{\rho}\, \frac{d\rho}{dr}\, 
   \xi^{(0)}_{\ell,m,k} - \frac{1}{\rho}\, \frac{\ell (\ell+1)}{r^2}\, F\, \xi^{(0)}_{\ell,m,k} } \nonumber \\
 & & \phantom{www}  + \frac{1}{\rho\, r^2}\, \frac{d}{dr} \left\{ r^2\, F \left[ \left( 3 - \kappa_T \right)\, 
   \frac{1}{T}\, \frac{dT}{dr}\, \xi^{(0)}_{\ell,m,k}  - \left( 1 + \kappa_\rho \right)\,  \frac{1}{\rho}\, 
   \frac{d\rho}{dr}\, \xi^{(0)}_{\ell,m,k} - \kappa_\mu\, \frac{1}{\mu}\, \frac{d\mu}{dr}\, \xi^{(0)}_
{\ell,m,k}   
   + \left( \frac{dT}{dr} \right)^{-1} \frac{d}{dr} \left( \frac{dT_{\ell,m,k} }{dr}\, \xi^{(0)}_{\ell,m,k} 
\right) \right] \right\}. 
   \hspace{1cm}
\end{eqnarray}
This expression can be simplified by noting that
\begin{equation}
\frac{\vec{\nabla} \cdot \vec{F}}{F} = \frac{2}{r} + \frac{3}{T}\, \frac{dT}{dr} 
   - \frac{1}{\rho}\, \frac{d\rho}{dr} - \frac{1}{\kappa}\, \frac{d\kappa}{dr} 
   + \left( \frac{dT}{dr} \right)^{-1} \frac{d^2 T}{dr^2}.
\end{equation}
After some algebra and using the property that $dQ/dt=0$ in the unperturbed star, it follows that
\begin{equation}
 \left( {{dQ} \over {dt}} \right)^{\! \prime (0)}_{\ell,m,k}
   = \frac{F}{\rho} \left[ \frac{d^2 \xi^{(0)}_{\ell,m,k}}{dr^2}
   + 2 \left( \frac{1}{F}\, \frac{dF}{dr} + \frac{1}{r} \right) \frac{d \xi^{(0)}_{\ell,m,k}}{dr}
   + \frac{\ell(\ell+1) - 2}{r^2}\, \xi^{(0)}_{\ell,m,k} \right].
\end{equation}
By using differential Eq.~(\ref{psi0}) and the Relation~(\ref{zero}) between $\Psi^{(0)}_{\ell,m,k}$ and $
\xi^{(0)}_{\ell,m,k}$, this expression can be further simplified to
\begin{equation}
\left( {{dQ} \over {dt}} \right)^{\! \prime (0)}_{\ell,m,k}  = 2\, \frac{F}{\rho} \left( \frac{1}{F}\, \frac
{dF}{dr} 
   - \frac{1}{g}\, \frac{dg}{dr} \right) \frac{d\xi^{(0)}_{\ell,m,k}}{dr}, 
\end{equation}
or, equivalently,
\begin{equation}
\left( {{dQ} \over {dt}} \right)^{\! \prime (0)}_{\ell,m,k}  = - \frac{L_r}{2\, \pi\, \rho\, g\, r^2} 
   \left( \frac{1}{L_r}\, \frac{dL_r}{dr} - \frac{\rho}{g} \right) \left[ \frac{d\Psi^{(0)}_{\ell,m,k}}{dr} 
   - \left( \frac{\rho}{g} - \frac{2}{r} \right) \Psi^{(0)}_{\ell,m,k} \right],  \label{dQdt0}
\end{equation}
where $L_r = 4\, \pi\, r^2\, F$ is the luminosity at radius $r$. Substitution of this expression into 
Eq.~(\ref{f2}) and differentiation of the radial component of the tidal displacement field with respect 
to time yields an expression for the radial component of the tidal velocity field in agreement with 
Eq.~(20) from Campbell (1984).

\section{C. Solution of the System of Differential Equations Governing Quasi-Static Tides}
\label{solution}

In the quasi-static approximation, the total perturbation of the gravitational potential associated with the 
spherical harmonic $Y_\ell^m(\theta,\phi)$ and forcing angular frequency $\sigma_{m,k}$ in Expansion~
(\ref{W}) of the tide-generating potential is given by
\begin{equation}
\Psi_{\ell,m,k}(r) = \Psi^{(0)}_{\ell,m,k}(r) + \sigma_{m,k}\, \Psi^{(1)}_{\ell,m,k}(r),
\end{equation}
where the functions $ \psi^{(0)}(r)$ and $ \psi^{(1)}(r)$ are determined by differential Eqs.~(\ref{psi0}) 
and~(\ref{psi1}), respectively. 
A general solution to Eq.~(\ref{psi0}) consists of a linear combination of two independent particular 
solutions $\psi_1^{(0)}(r)$ and $\psi_2^{(0)}(r)$:
\begin{equation}
\Psi^{(0)}_{\ell,m,k}(r) = C_1\, \psi_1^{(0)}(r) + C_2\, \psi_2^{(0)}(r), 
\end{equation}
where $C_1$ and $C_2$ are two undetermined constants. We let $\psi_1^{(0)}(r)$ be the particular solution 
that behaves as $r^\ell$ near $r = 0$, and $\psi_2^{(0)}(r)$ the particular solution that behaves as $r^{-
\ell-1}$ near $r = 0$. In order for the total perturbation of the gravitational potential and the radial 
component of the tidal displacement field to remain finite at $r=0$, we set $C_2=0$. The constant $C_1$ is 
determined by means of boundary Condition~(\ref{bcpsi0})\footnote{Note that we use the same dimensionless quantities as those adopted in \S\,\ref{qstide}.}:
\begin{equation}
C_1 = - \varepsilon_{\rm tide}\, (2\, \ell+1)\, c_{\ell,m,k} 
 \left[ \left( {{d\psi_1^{(0)}} \over {dr}} \right)_{r=1} 
  + \left( \ell + 1 - \frac{\rho_s}{g_s} \right)\, \psi_1^{(0)}(1)  \right]^{-1}, 
\end{equation}
where $\rho_s$ and $g_s$ are the mass density and gravity at the star's surface.

Next, a general solution to Eq.~(\ref{psi1}) is given by
\begin{equation}
\Psi^{(1)}_{\ell,m,k}(r) = C_1^\prime\, \psi_1^{(1)}(r) + C_2^\prime\, \psi_2^{(1)}(r) 
   + {\rm i}\, \psi_1^{(1)}(r) \int_0^r {{\beta(r)} \over {\Delta(r)}}\, \psi_2^{(1)}(r)\, dr
   - {\rm i}\, \psi_2^{(1)}(r) \int_0^r {{\beta(r)} \over {\Delta(r)}}\, \psi_1^{(1)}(r)\, dr, 
   \label{apppsi1}
\end{equation}
where $C_1^\prime$ and $C_2^\prime$ are two undetermined constants, $\psi_1^{(1)}(r)$ and $\psi_2^
{(1)}(r)$ are two independent particular solutions of the homogeneous part of Eq.~(\ref{psi1}), and the 
functions $\beta(r)$ and $\Delta(r)$ are defined as
\begin{equation}
\beta(r) = {1 \over {g\, r^2}}\, {d \over {dr}} \left[ \rho\, r^2\, \nu\, {d \over {dr}} \left( {{C_1\, \psi_1^{(0)}} 
\over g} \right) \right] 
\phantom{ww} \mbox{and} \phantom{ww}
\Delta(r) =  \psi_1^{(1)}\, {{d  \psi_2^{(1)}} \over {dr}} -  \psi_2^{(1)}\, {{d  \psi_1^{(1)}} \over {dr}}.
\end{equation}
The homogeneous part of Eq.~(\ref{psi1}) is formally the same as Eq.~(\ref{psi0}). We therefore adopt the 
same two independent particular solutions as before, and let $\psi_1^{(1)}(r)$ be the particular solution that 
behaves as $r^\ell$ near $r = 0$, and $\psi_2^{(1)}(r)$ the particular solution that behaves as $r^{-\ell-1}$ 
near $r = 0$. 

From a numerical point of view, it is convenient to avoid the numerical calculation of the derivatives of the 
mass density $\rho$ and turbulent viscosity coefficient $\nu$ in the function $\beta(r)$ appearing in the solution 
for $\psi^{(1)}(r)$. By performing an integration by parts and using that $\psi_1^{(1)}(r)$ and $\psi_2^
{(1)}(r)$ are solutions to the homogeneous part of Eq.~(\ref{psi1}), we therefore transform the solution for 
$\psi^{(1)}(r)$ into the form
\begin{equation}
\Psi^{(1)}_{\ell,m,k}(r) = C_1^\prime\, \psi_1^{(1)}(r)  + C_2^\prime\, \psi_2^{(1)}(r) 
  - {\rm i}\, \psi_1^{(1)}(r)\, \int_0^r \frac{\delta(r)}{\Delta(r)}\, \zeta_2(r)\, dr 
 + {\rm i}\, \psi_2^{(1)}(r)\, \int_0^r \frac{\delta(r)}{\Delta(r)}\, \zeta_1(r)\, dr, 
\end{equation}
where
\begin{equation}
\delta(r) = C_1\, {{\rho\, \nu} \over g^2} 
  \left[ {{d\psi_1^{(0)}} \over {dr}} - \left( {\rho \over g} - {2 \over r} \right) \psi_1^{(0)} \right]\!, 
  \phantom{wwww}
\zeta_1(r) = {{d\psi_1^{(1)}} \over {dr}} - \left( {\rho \over g} - {2 \over r} \right) \psi_1^{(1)}, 
  \phantom{ww} \mbox{and} \phantom{ww}
\zeta_2(r) = {{d\psi_2^{(1)}} \over {dr}} - \left( {\rho \over g} - {2 \over r} \right) \psi_2^{(1)}. 
\end{equation}
Because of the requirement that the total perturbation of the gravitational potential and the 
radial component of the tidal displacement field must remain finite at $r=0$, we set $C_2^\prime=0$. 
Boundary Condition~(\ref{bcpsi1}) then yields 
\begin{equation}
C_1^\prime = {\rm i}\, \frac{\chi_2(1) + (g_s/N_s^2) 
   \left[ \beta(1) + \lambda(1) \right]}{\chi_1(1)}, 
\end{equation}
where $N_s^2$ is the square of the Brunt-V\"{a}is\"{a}l\"{a} frequency at the surface of the unperturbed 
star, and
\begin{equation}
\lambda(r) = C^\prime (\Gamma_3-1)\, {{\rho} \over {c_s^2}}\, 
   \left( {{dQ} \over {dt}} \right)^{\! \prime (0)}, 
\end{equation}
\begin{equation}
\chi_1(r) = {{d\psi_1^{(1)}} \over {dr}} - \left( \ell +1 - {\rho \over g} \right) \psi_1^{(1)}, 
\end{equation}
\begin{eqnarray}
\lefteqn{ \chi_2(r) = \frac{d\psi_1^{(1)}}{dr} \int_0^r \frac{\delta(r)}{\Delta(r)}\, \zeta_2(r)\, dr
   - \frac{d\psi_2^{(1)}}{dr} \int_0^r \frac{\delta(r)}{\Delta(r)}\, \zeta_1(r)\, dr
   + \frac{\delta(r)}{\Delta(r)} \left[ \psi_1^{(1)}(r)\, \, \zeta_2(r)
   - \psi_2^{(1)}(r)\, \zeta_1(r) \right] } \nonumber \\
  & & \phantom{ww} 
   + \left( \ell + 1 - \frac{\rho}{g} \right) \left[
   \psi_1^{(1)}(r)\, \int_0^r \frac{\delta(r)}{\Delta(r)}\, \zeta_2(r)\, dr 
   - \psi_2^{(1)}(r)\, \int_0^r \frac{\delta(r)}{\Delta(r)}\, \zeta_1(r)\, dr \right].
\end{eqnarray}
The constant $C_1^\prime$ is thus purely imaginary.

\section{D. Alternate form of the tidal evolution equations}
\label{alt}

Equations~(\ref{da4}), (\ref{de4}), and~(\ref{domsec}) for the rates of secular change of the orbital 
semi-major axis $a$, the orbital eccentricity $e$, and the rotational angular velocity $\Omega_1$ are of a 
different form than the equations derived by, e.g., Zahn (1977, 1978), Hut (1981), and Ruymaekers (1992). 
To show that the different forms are equivalent, we restrict ourselves to the dominant $\ell=2$ terms and 
use Kepler's third law to rewrite Eq.~(\ref{da4}) as
\begin{equation}
\left( {{da} \over {dt}} \right)_{\rm sec} = 4\, \frac{G\,M_1}{R_1^3}\, q\, (1+q)  
  \left( {R_1 \over a} \right)^8 \frac{a}{n} 
  \sum_{m=-2}^2 \sum_{k=0}^{+\infty} \kappa_{2,m,k}\,
  \left| F_{2,m,k} \right|\, \sin \gamma_{2,m,k}\, G_{2,m,k}^{(2)}(e).  
  \label{altda}
\end{equation}
If we furthermore assume dissipative effects to be small, the phase angles $\gamma_{2,m,k}$ are small and 
proportional to the forcing angular frequencies $\sigma_{m,k}$, so that we can set
\begin{equation}
\sin \gamma_{2,m,k} \approx - \sigma_{m,k}\, \tau_2. \label{tau2}
\end{equation}
The constant $\tau_2$ has dimensions of time and is independent of $m$ and $k$ because the quasi-static tide 
solutions $\Psi_{\ell,m,k}^{(0)}(R_1)$ and $\Psi_{\ell,m,k}^{(1)}(R_1)$ are divided by $c_{\ell,m,k}$ in 
Eq.~(\ref{tang}) for the phase angles $\gamma_{\ell,m,k}$. The minus sign in Eq.~(\ref{tau2}) is included
to take into account that the tides lag behind the position of the companion when $\Omega_1 < n$.

By the use of the definition of the forcing angular frequencies $\sigma_{m,k}$, Eq.~(\ref{altda}) can 
then be cast in the form
\begin{equation}
\left( {{da} \over {dt}} \right)_{\rm sec} = - 12\, \frac{G\,M_1}{R_1^3}\, \tau_2\, q\, (1+q)  
  \left( {R_1 \over a} \right)^8 \frac{a}{\left( 1 - e^2 \right)^{15/2}}\,  
  \left[ f^{(1)}_2(e^2) - (1-e^2)^{3/2}\, f^{(2)}_2(e^2)\, \frac{\Omega_1}{n} \right], 
  \label{altda2}
\end{equation}
where
\begin{equation}
f^{(1)}_2(e^2) = \frac{1}{3} (1-e^2)^{15/2} \sum_{m=-2}^2 \sum_{k=0}^{+\infty}\,  
  k\, \kappa_{2,m,k}\, \left| F_{2,m,k} \right|\, G_{2,m,k}^{(2)}(e), 
\end{equation}
\begin{equation}
f^{(2)}_2(e^2) =  - \frac{1}{3} (1-e^2)^6 \sum_{m=-2}^2 \sum_{k=0}^{+\infty}\,  
  m\, \kappa_{2,m,k}\, \left| F_{2,m,k} \right|\, G_{2,m,k}^{(2)}(e). 
\end{equation}
In the limiting case where all forcing angular frequencies $\sigma_{m,k}$ tend to zero, the constants 
$F_{2,m,k}$ all tend to the classical apsidal motion constant $k_2$ (Smeyers \& Willems 2001). It can  
then be shown numerically or through the use of Taylor series that 
\begin{equation}
f^{(1)}_2(e^2)  \rightarrow k_2 \left( 1 + \frac{31}{2}\, e^2 + \frac{255}{8}\, e^4 
   + \frac{185}{16}\, e^6 + \frac{25}{64}\, e^8 \right),
\end{equation}
\begin{equation}
f^{(2)}_2(e^2)  \rightarrow k_2 \left( 1 + \frac{15}{2}\, e^2 + \frac{45}{8}\, e^4 
   + \frac{5}{16}\, e^6 \right).
\end{equation}
After setting $R_1^3/(G\, M_1\, \tau_2) =  t_F$, where $t_F$ is a characteristic tidal energy dissipation 
time scale, the equation for the rate of secular change of the orbital semi-major axis in the limiting case 
of weak damping and small forcing angular frequencies takes the same form as the equation for the rate 
of secular change of the orbital semi-major axis derived by Zahn (1977, 1978), Hut (1981), and 
Ruymaekers (1992)\footnote{Equation~(\ref{altda2}) differs from Eq.~(9) in Hut (1981) by a factor 
of 2 due to a different definition of the apsidal motion constant $k_2$.}.

Similarly, retaining only the dominant $\ell=2$ terms, Eqs.~(\ref{de4}) and~(\ref{domsec}) can be rewritten as
\begin{equation}
\left( {{de} \over {dt}} \right)_{\rm sec}  = -54\, \frac{G\,M_1}{R_1^3}\, \tau_2\, q\, (1+q)  
  \left( {R_1 \over a} \right)^8 \frac{e}{\left( 1 - e^2 \right)^{13/2}}\,  
  \left[ f^{(3)}_2(e^2) - \frac{11}{18} (1-e^2)^{3/2}\, f^{(4)}_2(e^2)\, 
  \frac{\Omega_1}{n} \right], \label{altde2}
\end{equation}
\begin{equation}
\left( {{d\Omega_1} \over {dt}} \right)_{\rm sec}  = 6\, \frac{G\,M_1}{R_1^3}\, \tau_2\, q^2\,   
  \frac{M_1\,R_1^2}{I_1}
  \left( {R_1 \over a} \right)^6 \frac{n}{\left( 1 - e^2 \right)^6}\,  
  \left[ f^{(5)}_2(e^2) - (1-e^2)^{3/2}\, f^{(6)}_2(e^2)\, 
  \frac{\Omega_1}{n} \right],  \label{altdo}
\end{equation}
where 
\begin{equation}
f^{(3)}_2(e^2) = \frac{2}{27} \frac{(1-e^2)^{13/2}}{e} \sum_{m=-2}^2 \sum_{k=0}^{+\infty}\,  
  k\, \kappa_{2,m,k}\, \left| F_{2,m,k} \right|\, G_{2,m,k}^{(3)}(e), 
\end{equation}
\begin{equation}
f^{(4)}_2(e^2) =  - \frac{4}{33} \frac{(1-e^2)^5}{e} \sum_{m=-2}^2 \sum_{k=0}^{+\infty}\,  
  m\, \kappa_{2,m,k}\, \left| F_{2,m,k} \right|\, G_{2,m,k}^{(3)}(e). 
\end{equation}
\begin{equation}
f^{(5)}_2(e^2) = - \frac{2}{3} (1-e^2)^6 \sum_{m=-2}^2 \sum_{k=0}^{+\infty}\,  
  k\, \kappa_{2,m,k}\, \left| F_{2,m,k} \right|\, G_{2,m,k}^{(4)}(e). 
\end{equation}
\begin{equation}
f^{(6)}_2(e^2) =  \frac{2}{3} (1-e^2)^{9/2} \sum_{m=-2}^2 \sum_{k=0}^{+\infty}\,  
  m\, \kappa_{2,m,k}\, \left| F_{2,m,k} \right|\, G_{2,m,k}^{(4)}(e). 
\end{equation}
In the limiting case where all forcing angular frequencies $\sigma_{m,k}$ tend to zero, it can again
be shown numerically or through the use of Taylor series that 
\begin{equation}
f^{(3)}_2(e^2)  \rightarrow k_2 \left( 1 + \frac{15}{4}\, e^2 + \frac{15}{8}\, e^4 
   + \frac{5}{64}\, e^6 \right),
\end{equation}
\begin{equation}
f^{(4)}_2(e^2)  \rightarrow k_2 \left( 1 + \frac{3}{2}\, e^2 + \frac{1}{8}\, e^4 
    \right),
\end{equation}
\begin{equation}
f^{(5)}_2(e^2)  \rightarrow k_2 \left( 1 + \frac{15}{2}\, e^2 + \frac{45}{8}\, e^4 
   + \frac{5}{16}\, e^6 \right).
\end{equation}
\begin{equation}
f^{(6)}_2(e^2)  \rightarrow k_2 \left( 1 + 3\, e^2 + \frac{3}{8}\, e^4 
   \right).
\end{equation}
Hence, after setting $R_1^3/(G\, M_1\, \tau_2) =  t_F$ the equations for the rate of secular change of the 
orbital eccentricity and rotational angular velocity in the limiting case of weak damping and small forcing 
angular frequencies also take the same form as the equations for the rate of secular change of the orbital 
eccentricity and rotational angular velocity derived by Zahn (1977, 1978), Hut (1981), and Ruymaekers (1992).


\begin{thebibliography}{}
\bibitem[Aizenman \& Smeyers(1977)]{1977Ap&SS..48..123A} Aizenman, M.~L., \& Smeyers, P.\ 1977, 
\apss, 48, 123 
\bibitem[Alexander(1973)]{1973Ap&SS..23..459A} Alexander, M.~E.\ 1973, \apss, 23, 459 
\bibitem[Belczynski et al.(2008)]{2008ApJS..174..223B} Belczynski, K., Kalogera, V., Rasio, F.~A., 
Taam, R.~E., Zezas, A., Bulik, T., Maccarone, T.~J., \& Ivanova, N.\ 2008, \apjs, 174, 223 
\bibitem[Bender et al.(1998)]{Bender} Bender,~P. et al.\ 1998 LISA Pre-Phase A Report, 2nd Ed. (MPQ)
\bibitem[Brouwer \& Clemence(1961)]{1961mcm..book.....B} Brouwer, D., \& Clemence, G.~M.\ 1961, 
New York: Academic Press 
\bibitem[Campbell(1984)]{1984MNRAS.207..433C} Campbell, C.~G.\ 1984, \mnras, 207, 433 
\bibitem[Cox \& Giuli(1968)]{1968QB801.C65......} Cox, J.~P., \& Giuli, R.~T.\ 1968, New York, 
Gordon and Breach
\bibitem[Deloye \& Taam(2006)]{2006ApJ...649L..99D} Deloye, C.~J., \& Taam, R.~E.\ 2006, \apjl, 649, L99 
\bibitem[Deloye et al.(2007)]{2007MNRAS.381..525D} Deloye, C.~J., Taam, R.~E., Winisdoerffer, C., \& Chabrier, G.\ 2007, \mnras, 381, 525
\bibitem[Dziembowski(1982)]{1982AcA....32..147D} Dziembowski, W.\ 1982, Acta Astronomica, 32, 147 
\bibitem[Eggleton(1983)]{1983ApJ...268..368E} Eggleton, P.~P.\ 1983, \apj, 268, 368
\bibitem[Fitzpatrick(1970)]{1970QB351.F56......} Fitzpatrick, P.~M.\ 1970, New York, Academic Press
\bibitem[Gokhale et al.(2007)]{2007ApJ...655.1010G} Gokhale, V., Peng, X.~M., \& Frank, J.\ 2007, \apj, 
655, 1010 
\bibitem[Goldman \& Mazeh(1991)]{1991ApJ...376..260G} Goldman, I., \& Mazeh, T.\ 1991, \apj, 376, 260 
\bibitem[Goldreich \& Keeley(1977)]{1977ApJ...211..934G} Goldreich, P., \& Keeley, D.~A.\ 1977, \apj, 211, 934
\bibitem[Goodman \& Oh(1997)]{1997ApJ...486..403G} Goodman, J., \& Oh, S.~P.\ 1997, \apj, 486, 403  
\bibitem[Hansen et al.(2004)]{2004sipp.book.....H} Hansen, C.~J., Kawaler, S.~D., \& Trimble, V.\ 2004, 
Stellar interiors : physical principles, structure, and evolution, 2nd ed., New York: Springer-Verlag  
\bibitem[Hut(1981)]{1981A&A....99..126H} Hut, P.\ 1981, \aap, 99, 126 
\bibitem[Iben et al.(1998)]{1998ApJ...503..344I} Iben, I.~J., Tutukov, A.~V., \& Fedorova, A.~V.\ 1998, \apj, 503, 344 
\bibitem[Ledoux \& Walraven(1958)]{1958HDP....51..353L} Ledoux, P., \& Walraven, T.\ 1958, 
Handbuch der Physik, 51, 353 
\bibitem[Marsh et al.(2004)]{2004MNRAS.350..113M} Marsh, T.~R., Nelemans, G., \& Steeghs, D.\ 2004, 
\mnras, 350, 113
\bibitem[Mathieu et al.(2004)]{2004ApJ...602L.121M} Mathieu, R.~D., Meibom,  S., \& Dolan, C.~J.\ 
2004, \apjl, 602, L121 
\bibitem[Meibom \& Mathieu(2005)]{2005ApJ...620..970M} Meibom, S., \& Mathieu, R.~D.\ 2005, \apj, 
620, 970 
\bibitem[Menou et al.(2008)]{2008NewAR..51..884M} Menou, K., Haiman, Z., \& Kocsis, B.\ 2008, New Astronomy Review, 51, 884 
\bibitem[Mochkovitch \& Livio(1989)]{1989A&A...209..111M} Mochkovitch, R., \& Livio, M.\ 1989, \aap, 209, 111 
\bibitem[Nelemans et al.(2001)]{2001A&A...365..491N} Nelemans, G., Yungelson, L.~R., Portegies 
Zwart, S.~F., \& Verbunt, F.\ 2001, \aap, 365, 491
\bibitem[Osaki \& Hansen(1973)]{1973ApJ...185..277O} Osaki, Y., \& Hansen, C.~J.\ 1973, \apj, 185, 277 
\bibitem[Penev et al.(2007)]{2007ApJ...655.1166P} Penev, K., Sasselov, D., Robinson, F., \& Demarque, P.\ 2007, \apj, 655, 1166 
\bibitem[Peters(1964)]{1964PhRv..136.1224P} Peters, P.~C.\ 1964, Physical Review , 136, 1224 
\bibitem[Polfliet \& Smeyers(1990)]{1990A&A...237..110P} Polfliet, R., \& Smeyers, P.\ 1990, \aap, 237, 
110 
\bibitem[Racine et al.(2007)]{2007MNRAS.380..381R} Racine, {\'E}., Phinney, E.~S., \& Arras, P.\ 2007, \mnras, 380, 381
\bibitem[Rathore et al.(2005)]{2005MNRAS.357..834R} Rathore, Y., Blandford, R.~D., \& Broderick, A.~E.\ 2005, \mnras, 357, 834 
\bibitem[Ruymaekers(1992)]{1992A&A...259..349R} Ruymaekers, E.\ 1992, \aap, 259, 349  
\bibitem[Savonije \& Papaloizou(1983)]{1983MNRAS.203..581S} Savonije, G.~J., \& Papaloizou, 
J.~C.~B.\ 1983, \mnras, 203, 581 
\bibitem[Savonije \& Papaloizou(1984)]{1984MNRAS.207..685S} Savonije, G.~J., \& Papaloizou, 
J.~C.~B.\ 1984, \mnras, 207, 685 
\bibitem[Savonije \& Witte(2002)]{2002A&A...386..211S} Savonije, G.~J., \& Witte, M.~G.\ 2002, \aap, 
386, 211
\bibitem[Sesana et al.(2008)]{2008MNRAS.391..718S} Sesana, A., Vecchio, A., Eracleous, M., \& Sigurdsson, S.\ 2008, \mnras, 391, 718 
\bibitem[Smeyers(1997)]{1997A&A...318..140S} Smeyers, P.\ 1997, \aap, 318, 140 
\bibitem[Smeyers et al.(1991)]{1991A&A...248...94S} Smeyers, P., van Hout, M., Ruymaekers, E., \& 
Polfliet, R.\ 1991, \aap, 248, 94 
\bibitem[Smeyers et al.(1998)]{1998A&A...335..622S} Smeyers, P., Willems, B., \& Van Hoolst, T.\ 1998, 
\aap, 335, 622 
\bibitem[Smeyers \& Willems(1998)]{1998A&A...336..539S} Smeyers, P., \& Willems, B.\ 1998, \aap, 
336, 539 
\bibitem[Smeyers \& Willems(2001)]{2001A&A...373..173S} Smeyers, P., \& Willems, B.\ 2001, \aap, 
373, 173 
\bibitem[Sterne(1939)]{1939MNRAS..99..451S} Sterne, T.~E.\ 1939, \mnras, 99, 451 
\bibitem[Sterne(1960)]{1960aitc.book.....S} Sterne, T.~E.\ 1960, Interscience Tracts on Physics and 
Astronomy, New York: Interscience Publication
\bibitem[Terquem et al.(1998)]{1998ApJ...502..788T} Terquem, C., Papaloizou, J.~C.~B., Nelson, R.~P., 
\& Lin, D.~N.~C.\ 1998, \apj, 502, 788 
\bibitem[Unno et al.(1989)]{1989nos..book.....U} Unno, W., Osaki, Y., Ando, H., Saio, H., \& Shibahashi, 
H.\ 1989, Nonradial oscillations of stars, Tokyo: University of Tokyo Press, 1989, 2nd ed.
\bibitem[Van Hoolst(1994a)]{1994A&A...286..879V} Van Hoolst, T.\ 1994a, \aap, 286, 879 
\bibitem[Van Hoolst(1994b)]{1994A&A...292..471V} Van Hoolst, T.\ 1994b, \aap, 292, 471 
\bibitem[Willems(2000)]{2000PhDT.........2W} Willems, B.\ 2000, Ph.D.~Thesis, Katholieke Universiteit 
Leuven, Belgium
\bibitem[Willems(2003)]{2003MNRAS.346..968W} Willems, B.\ 2003, \mnras, 346, 968 
\bibitem[Willems et al.(2003)]{2003A&A...397..973W} Willems, B., van Hoolst, T., \& Smeyers, P.\ 
2003, \aap, 397, 973 
\bibitem[Willems et al.(2007)]{2007ApJ...665L..59W} Willems, B., Kalogera, V., Vecchio, A., Ivanova, 
N., Rasio, F.~A., Fregeau, J.~M., \& Belczynski, K.\ 2007, \apjl, 665, L59
\bibitem[Witte \& Savonije(1999)]{1999A&A...341..842W} Witte, M.~G., \& Savonije, G.~J.\ 1999, \aap, 
341, 842 
\bibitem[Witte \& Savonije(1999)]{1999A&A...350..129W} Witte, M.~G., \& Savonije, G.~J.\ 1999, \aap, 
350, 129 
\bibitem[Witte \& Savonije(2001)]{2001A&A...366..840W} Witte, M.~G., \& Savonije, G.~J.\ 2001, \aap, 
366, 840 
\bibitem[Witte \& Savonije(2002)]{2002A&A...386..222W} Witte, M.~G., \& Savonije, G.~J.\ 2002, \aap, 
386, 222 
\bibitem[Wu \& Goldreich(2001)]{2001ApJ...546..469W} Wu, Y., \& Goldreich, P.\ 2001, \apj, 546, 469 
\bibitem[Zahn(1966)]{1966AnAp...29..489Z} Zahn, J.~P.\ 1966, Annales d'Astrophysique, 29, 489 
\bibitem[Zahn(1975)]{1975A&A....41..329Z} Zahn, J.-P.\ 1975, \aap, 41, 329 
\bibitem[Zahn(1977)]{1977A&A....57..383Z} Zahn, J.-P.\ 1977, \aap, 57, 383 
\bibitem[Zahn(1978)]{1978A&A....67..162Z} Zahn, J.-R.\ 1978, \aap, 67, 162 
\bibitem[Zahn(2008)]{2008EAS....29...67Z} Zahn, J.-P.\ 2008, EAS Publications Series, 29, 67
\end{thebibliography}
\end{document}